\documentclass[aps,pre,twocolumn]{revtex4}
\usepackage{amsmath,bm,epsfig,,subfigure}
\usepackage{color}
\usepackage{epstopdf}
\usepackage[normalem]{ulem}
\def\Fbox#1{\vskip1ex\hbox to 8.5cm{\hfil\fboxsep0.3cm\fbox{%
  \parbox{8.0cm}{#1}}\hfil}\vskip1ex\noindent}  

\newcommand{\B}[1]{{\bm{#1}}}
\newcommand{\C}[1]{{\mathcal{#1}}}    
\let \= \equiv \let\*\cdot \let\~\widetilde \let\-\overline

\begin{document}
\title{Elasticity and Plasticity in Stiff and Flexible Oligomeric Glasses}
\author{Oleg Gendelman, H. George E. Hentschel$^*$, Pankaj K. Mishra, Itamar Procaccia and Jacques Zylberg}
\affiliation{Department of Chemical Physics, The Weizmann Institute of Science, Rehovot 76100, Israel\\
$^*$ Department of Physics, Emory University, Atlanta GA, USA}
\date{\today}
\begin{abstract}
In this paper we focus on the mechanical properties of oligomeric glasses (waxes), employing a microscopic model
that provides, via numerical simulations,  information about the shear modulus of such materials, the failure mechanism via plastic instabilities and about the geometric responses of the oligomers themselves to a mechanical load. We present a microscopic theory
that explains the numerically observed phenomena, including an exact theory of the shear modulus and of the plastic instabilities,
both local and system spanning. In addition we present a model to explain the geometric changes in the oligomeric chains
under increasing strains.
\end{abstract}
\maketitle

\section{Introduction}

A polymer is a macromolecule that consists of a large number of monomer subunits~\cite{RC03}. Polymeric glasses are solids composed of a large number of such polymeric units. Subjected to homogeneous strain such solids can exhibit a variety of interesting phenomena including crazing instabilities, shear banding, strain hardening etc~\cite{DK82}. Considerable effort was expended to describe
these phenomena on the microscopic level using theory and simulations \cite{MAS93,WR07,VBB04}.
Under tensile strains cavities may nucleate in a hitherto homogeneous polymeric glass. It was argued that the formation of cavities takes place  in regions of local low elastic modulus~\cite{MPRLB11}. Polymeric glasses subjected to large strains exhibit strain hardening; this may suppress strain localization and consequent crazing, necking, shear banding etc. Strain hardening is presumably caused by ordering the polymer beyond a certain strain threshold. The microscopic origin of strain hardening was
studied using molecular dynamic simulations in Ref.~\cite{HR07,BBL10}, finding that the origin of this phenomenon is related
to plastic rearrangements of the monomers. This also leads to short-range ordering. In spite of the above mentioned efforts a first-principles theory of these interesting phenomena is still incomplete. In particular in this paper we propose a microscopic
theory that relates macroscopic observables with the conformational deformation of the oligomers under pure shear.

 In recent years there has been great progress in understanding the mechanical properties of amorphous solids
from first principles~\cite{ML99,ML06, LP09}. This progress was based on identifying elementary plastic events as the loss of
mechanical stability when a Hessian eigenvalue hits zero~\cite{ML99,ML06, LP09}. This event is connected to a saddle node
bifurcation in the generalized energy landscape. It was demonstrated also that these elementary events
can aggregate and concatenate to yield shear localization and eventually shear bands~\cite{DHP12,DGMPS13}.
The aim of this paper is to extend this analytic approach to plasticity from simple Lennard-Jones glasses (and recently some glasses with magnetic properties) \cite{12HIP,13DHPS,14HIPS} to the realm of short oligomeric (or wax)
glasses. These are amorphous solids whose constituents are short chains of the order of 10-30 monomers,
where the full impact of polymeric entanglement is still not crucial~\cite{BBP03}. Nevertheless the existence of fairly
long chains of connected monomers introduces a hierarchy of new length scales and energy scales related
to valence bonds, valence angles and inter-oligomer interactions. In particular the persistence length $\ell_p$ of the oligomer turns out to be crucial. Thus a variety of new phenomena and questions arise, calling for a careful numerical simulation and analytic assessment.
Among the issues arising we will provide a microscopic theory for the shear modulus of these materials, for the failure mechanism
through plasticity (both local and system spanning) and shed light on the geometric characteristic of the oligomers under
mechanical yield.

The outline of the paper is as follows:
In Sect. \ref{model} we describe the atomistic model used in further simulations. The model employs
Lennard-Jones, angular and FENE interactions (and see below for details). Sect. \ref{sim} presents firstly
the results of numerical simulations for the stress vs. strain curves, the energy budget, characteristic
of the oligomeric chains like end-to-end distance etc. For analytic transparency we perform the simulation
in quasi-static athermal conditions to highlight the plastic events without any thermal fluctuations or
strain rate effects that mask the fundamental physics. The same section provides some theory of these characteristics.  In Sect. \ref{plastic} we present a theory for
elementary plastic events. Next in Sect. \ref{failure} we discuss the failure mechanism involving shear localization and eventually shear bands. The following section \ref{modulus} presents the analytic calculation of the shear
modulus and a comparison with the numerics.

\section{Description of the model}
\label{model}
We consider a system composed of $N_p$ chains each comprising $n$ monomers (oligomers). Thus
the total number of particles in our system is $N=N_p\times n$. The interaction between monomers
belonging to the same or to different oligomers is different. Inter-oligomer interaction are simply
given by a truncated and smoothed Lennard-Jones potential $\phi_{\rm LJ}$, see below in
Eq. (\ref{LJpotential}). Within a given oligomer the interactions have three contributors. First, all monomers within the Lennard-Jones cutoff range $r_{\rm co}$ exert a force on each other which is derived from the potential $\phi_{\rm LJ}$. Secondly, a contribution $\chi$ is added to the energy of any two successive monomers within the polymer (to mimic
 the valence bond interaction). The third contribution to the energy is an angular potential to constrain the value of the valence angle $\theta$ determined by three successive monomers within a oligomer. This interaction is denoted below $\psi(\theta)$. Thus the total energy can be written as~\cite{KG89}
\begin{eqnarray}
U&=&U^{\rm LJ}+U^{\rm FENE}+U^{\rm Angle} \\
U^{\rm LJ}&=&\sum\limits_{\langle ij \rangle}^{N}\phi^{ij}_{\rm LJ} \ , \quad
U^{\rm FENE}=   \sum_{k=1}^{N_p}\sum\limits_{i=1}^{n-1}\chi_k^{i,i+1} \nonumber\\
U^{\rm Angle}&=&  \sum_{k=1}^{N_p} \sum\limits_{i=2}^{n-1}\psi_k^{i-1,i,i+1}
\end{eqnarray}
The notation is such that successive particles are $i$ and $i+1$ within a oligomer chain and $\psi_k^{i}$ stands for the angular contribution formed by any three successive particles $(i-1,i,i+1)$ within the $k'th$ oligomer where $i$ is the vertex.

The truncated and smoothed potential Lennard-Jones potential is defined as:
\begin{eqnarray}\label{LJpotential}
\phi^{ij}_{\rm LJ}& =&
4\varepsilon\left[\left(\frac{\lambda}{ r_{ij}}\right)^{12}-\left(\frac{\lambda}{r_{ij}}\right)^{6}\right]
\ , r_{ij}\le r_{\rm min}\\
\phi^{ij}_{\rm LJ}&=&\varepsilon\!\!\left[a\left(\frac{\lambda}{ r_{ij}}\right)^{12}\!\!\!\!\!\!\!-\!b\left(\frac{\lambda}{ r_{ij}}\right)^{6}
\!\!\!+ \!\!\!\sum_{\ell=0}^{3}c_{2\ell}\left(\!\frac{r_{ij}}{\lambda}\!\!\right)^{2\ell}\right] \nonumber\\
&~& \hskip 3 cm r_{\rm min}<r_{ij}<r_{\rm co} \ ,\\
\phi^{ij}_{\rm LJ}&=& 0
   \ , r_{ij}\ge r_{\rm co} \ .
\end{eqnarray}
Here $r_{\rm min}/\lambda$ is the length where the potential attains its minimum, and $r_{\rm co}/\lambda$
is the cut-off length for which the potential vanishes. The coefficients $a,~b$ and $c_{2\ell}$ are chosen such
that the repulsive and attractive parts of the potential are continuous with two derivatives at the potential minimum and the potential goes
to zero continuously at $r_{\rm co}/\lambda$ with two continuous derivatives as well.
The unit of length $\lambda=1.0$ is set to be the interaction length scale of two particles, $\varepsilon$ is the unit of energy and the Boltzmann constant $k_B = 1$.

For any two successive particles within the $k'th$ chain there is the Finite Non-Elastic Elongation (FENE) potential with finite length $r_0$ which is defined as:
\begin{equation}
\chi_k^{i,i+1}(r)=
\begin{cases}
 -\frac{1}{2}\eta r^2_0\ln[1-(r/r_0)^2] &; r<r_0 \\
\infty &; r\ge r_0
\end{cases}
\label{fene}
\end{equation}
where $r\equiv r_{i,i+1}/\lambda$ and $\eta$ is a parameter with units of force per unit length.

Finally, for any three successive monomers within the $k'th$ oligomer with vertex $i$ there is an angle constraint around a chosen equilibrium angle $\varphi^{\rm eq}$ and is defined as:
\begin{eqnarray}
\psi_k^{i-1,i,i+1}(\varphi^{i})&=&\kappa[\cos\varphi_k^{i}-\cos\varphi^{\rm eq}]^\alpha \nonumber\\
&=&\kappa[1+\cos\varphi_k^i]^\alpha\ .
\label{Angle_Potential}
\end{eqnarray}
Below we will also employ the angle $\theta$ were $\theta\equiv \pi-\varphi$. Thus for a stiff polymer $\varphi\approx \pi$
while $\theta$ is close to zero.

We distinguish between two cases, that of a stiff oligomer with $\alpha=1$
and a semi-flexible oligomer with $\alpha=2$. The meaning of the words ``stiff" and ``semi-flexible"
will be made clear in the sequel. The values of all the parameters used in the simulation are given
in Table \ref{table:par}.
\begin{table*}
\caption{The parameters used in the simulation.}
\begin{center}
 \begin{tabular}{||c c c c c c c c c||}
 \hline
a&b&c0&c2&c4&c6&$\eta$&$r_0$&$\varphi^{\rm eq}$\\
  \hline
  3.9435&-3.89268&1.2$\times10^{-3}$&-0.0207&0.10691&-0.143794&30&1.5&$\pi$\\
  \hline
 \end{tabular}

\end{center}
\label{table:par}
\end{table*}
\section{Numerical Simulations}
\label{sim}
We prepare a 2-dimensional system consisting of 256 polymers having 20 monomers in a chain. The initial density $\rho=0.8$ and the temperature is chosen such that the system is in the liquid state with high temperature T=1.3. To achieve such a state we begin with the crystalline arrangement of the polymers on
a square lattice, and we allow the crystal to melt by molecular dynamics. The masses of the monomers are all unity. The melt is equilibrated using a standard NPT procedure for 25~$\tau_{\alpha}$ LJ time units at pressure P=1.0 (LJ units), where $\tau_{\alpha}$ is the alpha relaxation time. After equilibration the polymer melt is coupled to a heat-bath at temperature T=0.01 (LJ units) and constant pressure (P=1). The system is then further equilibrated for another 100 LJ time units. Finally the glass sample is taken to the nearest inherent minimum state using a conjugate gradient scheme. This protocol is referred to as ``infinitely fast'' quench. Below we also consider samples prepared by finite quench rates.

Having prepared the oligomeric glass sample it is subjected to an Athermal Quasi-Static strain (AQS) as described in detail in ref.~\cite{LP09}. In brief, each monomer is first displaced by the affine transformation
\begin{equation}
 x_i\rightarrow x_i+\delta \gamma y_i; \hspace{0.5cm} y_i\rightarrow y_i,
\end{equation}
where $r_i\equiv(x_i,y_i)$ is the initial position of the $i^{\rm th}$ monomer and $\delta{\gamma}$ is the strain step applied during each affine transformation. The above transformation leads to non-zero resultant forces on the monomers. These forces are annulled by a non-affine transformation $r_i\rightarrow r_i+u_i$, where $u_i$ is displacement of the monomer necessary to return to mechanical equilibrium. The non-affine displacement is computed using the conjugate gradient scheme. The strain step is chosen for the present study is $\delta\gamma=10^{-4}$. The simulation is performed under periodic boundary condition along each direction of the box using the Lees-Edward formalism.

We consider both stiff and semi-flexible polymers in our studies (see Eq. (\ref{Angle_Potential})). For stiff polymers the results will be presented for $\kappa=2,5, 10,$ and 15;  for the semi-flexible case we consider $\kappa=2, 4, 8$, 10. Unless stated specifically the
results reported below will refer to the stiff oligomer case with $\alpha=1$.
\begin{figure}
\includegraphics[scale=0.5]{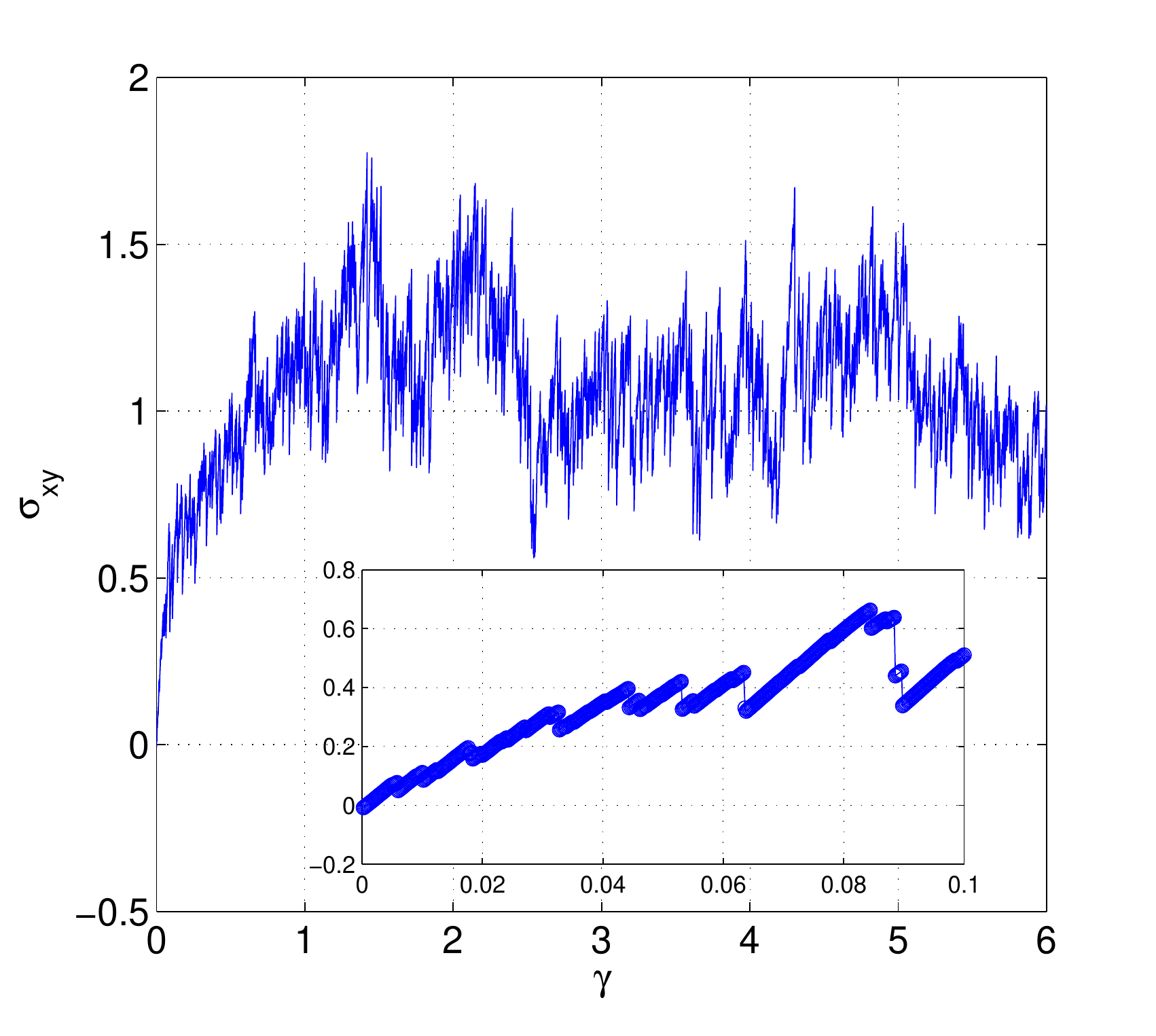}
\caption{A typical stress vs strain curve obtained by AQS straining of 256 polymers of chain length 20 with stiffness parameter $\kappa=2$. Smooth (linear) increases in the strain are punctuated with sharp drops.
The trajectory of stress vs strain is reversible only until the first drop. The sharp drops are plastic events as explained in the text. Inset: a blow-up of the first few plastic drops.}
\label{fig1}
\end{figure}

\subsection{Mechanical response of the polymer}
A typical stress vs. strain curve that results in the AQS protocol is shown in Fig.~\ref{fig1} for a single realization of a stiff oligomeric glass with $\kappa=2$. The stress grows linearly at first with the strain, and the protocol can be reversed to return to initial state. Upon increasing the strain the stress vs strain trajectory gets punctuated with sharp drops, these are irreversible, and after the occurrence of the first one we cannot return to the initial state by reversing the protocol.  After each plastic drop the stress rises again linearly with the applied strain (but not necessarily with the same slope) until the next plastic drop takes place. Generally speaking both the stress vs. strain and the energy vs. strain curves reach eventually a kind of steady state
in which the average stress and energy do no longer change even though they still experience elastic increases
and plastic drops.

\begin{figure}
\includegraphics[scale=0.45]{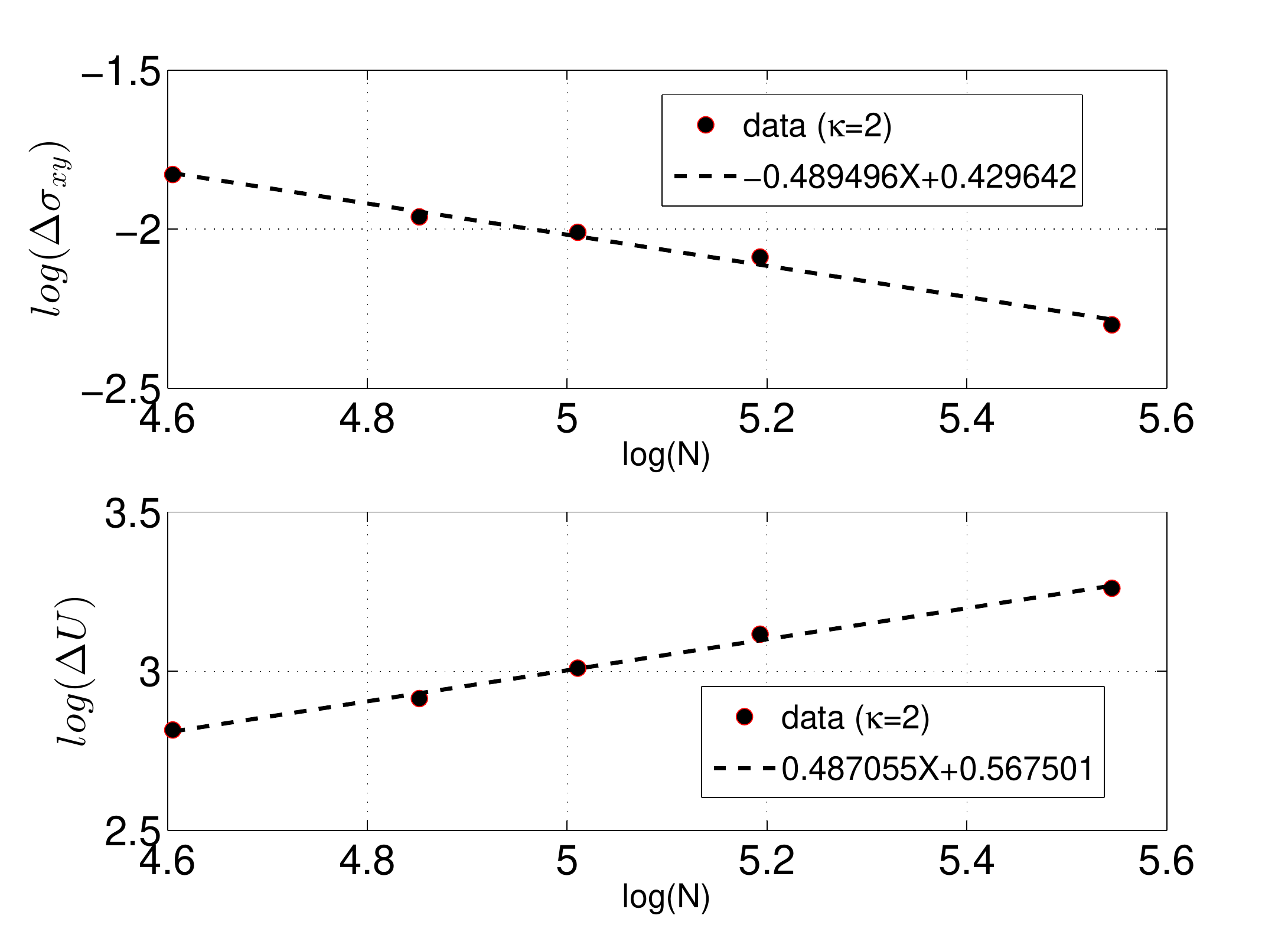}
 \caption{log-log plots of the average magnitude of the stress drops $\Delta \sigma$ and energy drops
 $\Delta U$ in the steady state (after the yield strain had been passed) as a function of the number of particles in the system}
 \label{average}
 \end{figure}

At first, when the external strain is still small, the energy drops associated with the plastic events are small, and
do not increase with the system size. These plastic energy drops are associated with localized events as is
explained in the next section. On the other hand, when the external strain is increased, at a threshold value of the external strain (also known as the yield strain $\sigma_{_{\rm Y}}$) much bigger energy drops become possible.
Once the yield stress has been achieved, there is a quantitative change in the nature of the plastic drops
since they become system-size dependent. We can examine the statistics of the magnitude of the energy and stress drops in the steady state. In Fig. \ref{average} we show the average magnitude of energy $\langle \Delta U \rangle$ and stress drops $\langle \Delta \sigma \rangle$ for systems of increasing number of particles $N$. It appears that the data support the scaling laws
\begin{eqnarray}
\langle \Delta U \rangle \sim N^\alpha\ ,  \quad \alpha\approx 0.5 \ , \label{alpha}\\
\langle \Delta \sigma \rangle \sim N^\beta \ , \quad \beta\approx -0.5 \ .\label{beta}
\end{eqnarray}
In Ref. \cite{LP09} it was shown that these exponents satisfy a scaling relation $\alpha-\beta=1$
as these exponents do. Since the system spanning events are confined to linear structures one is not surprised
with the exponent $\alpha=1/2$ in a 2-dimensional system. The scaling relation immediately determines also
$\beta=-1/2$.

\begin{figure}
\includegraphics[scale=0.45]{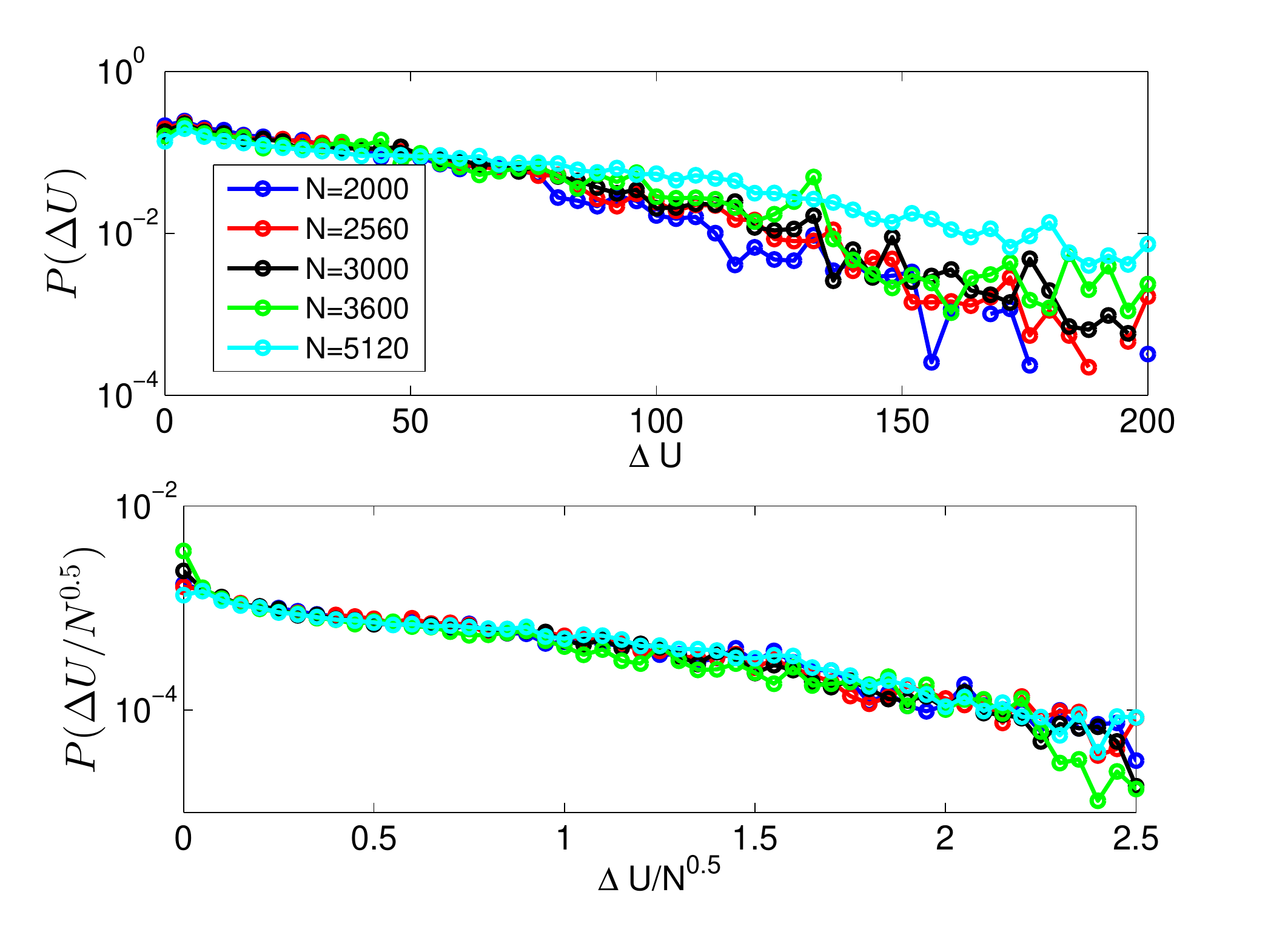}
 \caption{Raw and re-scaled pdf's for the energy drops in the steady state. The data collapse in the tails of the distributions supports the scaling laws presented in Eqs. (\ref{alpha}) and (\ref{beta}).}
 \label{scalstat}
 \end{figure}

To make sure that the exponent $\alpha=0.5$ is consistent we can test the probability distribution
functions (pdf) of the energy or stress drops. In Fig. \ref{scalstat} we show the raw pdfs of these
quantities and the re-scaled pdf's. The re-scaling is done using the exponent $\alpha=1/2$. The
data collapse of the pdfs in the tails shows that the exponent is adequate. Note that the re-scaling
does not collapse the data for small drops, these continue to be system size independent.

A theoretical discussion of the localized and the subextensive plastic events is provided in Sect. \ref{plastic}. Nevertheless the
reader should note that a continuum description of the stress vs strain curves in our open system is still under debate, even in the case of simpler examples like binary Lennard-Jones glasses. Here the quantitative theory of energy input by mechanical strain, including the share taken
by stress vs. oligomeric conformation changes on the one hand, and energy dissipated to the heat bath on the other hand is still unavailable.
Such an understanding is prerequisite to any continuum theory.

\subsection{Stress and energy averaged over realizations as a function of the stiffness parameter}

\begin{figure}
\includegraphics[scale=0.42]{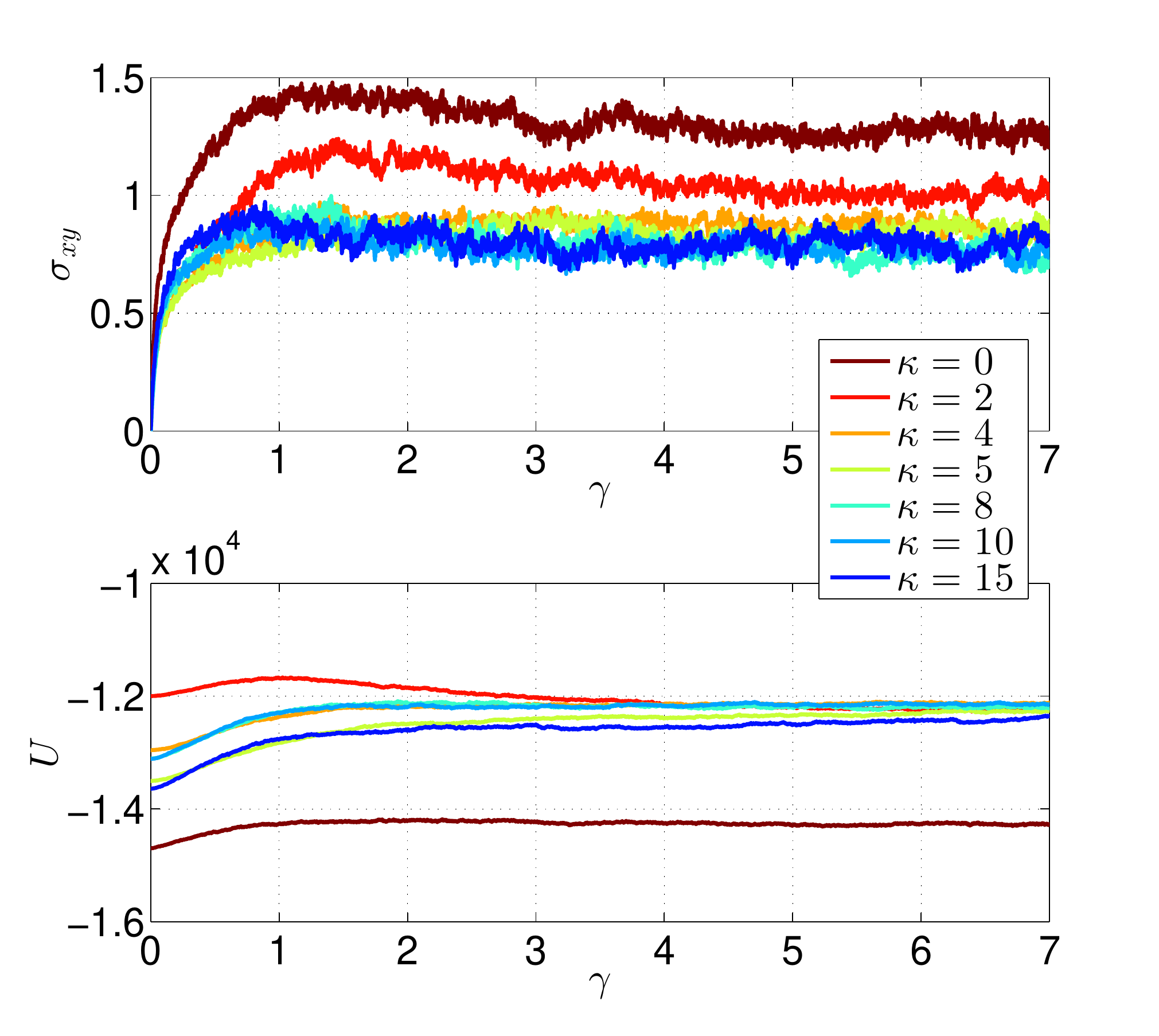}
 \caption{Upper panel: Stress Vs strain for different $\kappa$ for the stiff polymer ($\alpha=1$). The stress reaches to the same steady state for the finite $\kappa$. Bottom Panel: the variation of total internal potential energy with strain. The data averaged over 40 independent realizations are shown.}
 \label{fig4}
\end{figure}

\begin{figure}
 \includegraphics[scale=0.45]{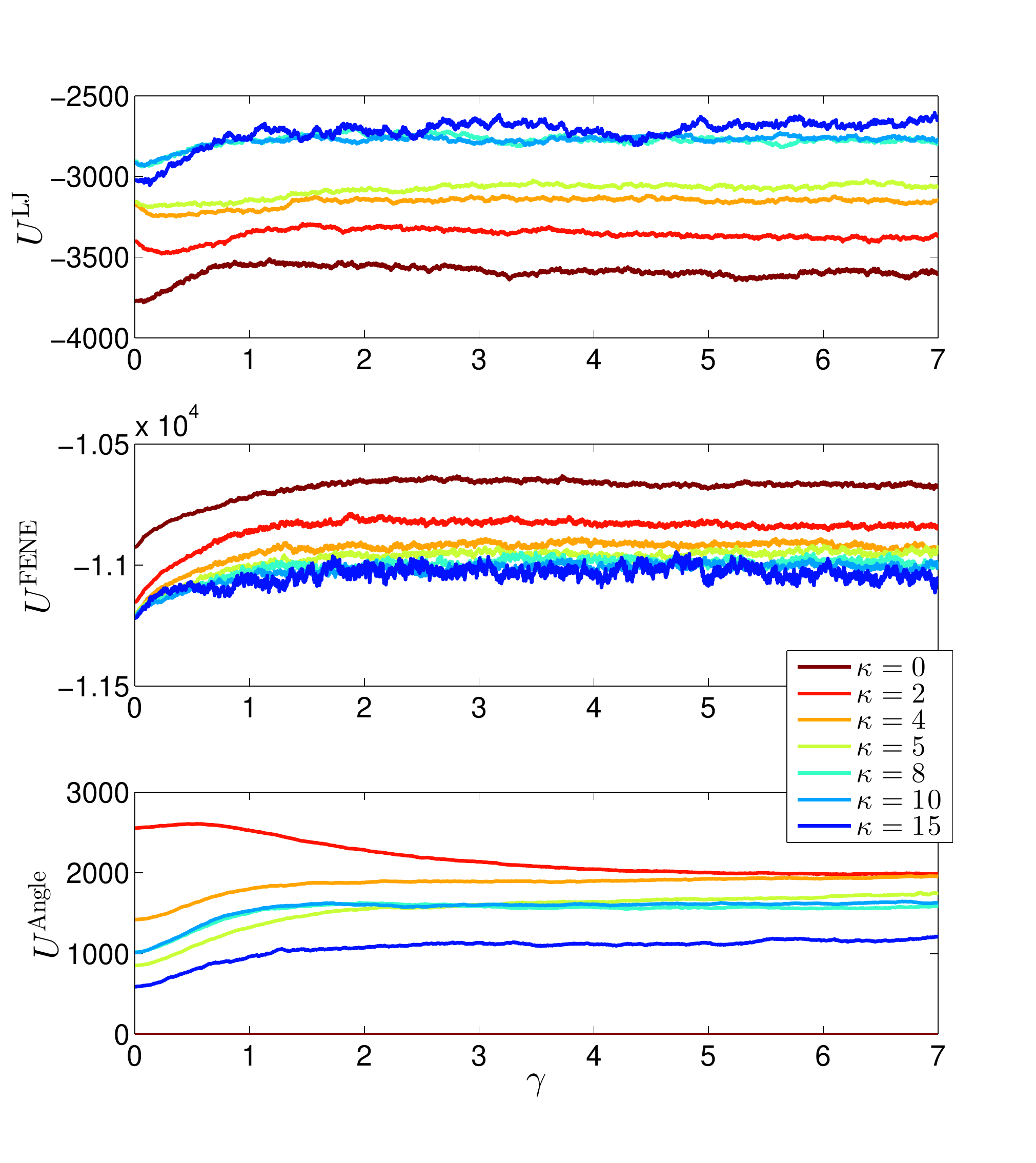}
 \caption{Variation of contribution in the total potential energy due to different interaction with strain. Upper panel: $U_{LJ}$ Vs strain for different $\kappa$. The potential energy due to the LJ interaction increases on increase of $\kappa$. Middle panel: $U_{FENE}$ Vs strain for different $\kappa$. Bottom panel: $U_{angle}$ Vs strain. The potential energy $U_{angle}$ decreases on increase of the stiffness of the chain.}
 \label{fig5}
\end{figure}

In addition to the measuring the plastic drops in single realization of the glass, it is interesting to examine
the energy and the stress averaged over many realizations. Such graphs should be closer to what is expected in the
thermodynamic limit when $N_p\to \infty$. In particular we can examine the dependence on the stiffness parameter
$\kappa$. In Fig. \ref{fig4} we see the stress vs. strain and the energy vs. strain averaged over 40 independent realizations as a function of $\kappa$.
It is interesting to see that both the energy and the stress appear to reach the same steady state for all $\kappa\ne 0$, but not for $\kappa=0$. To underline the fact that the attainment of the same
steady state is not at all trivial, we show in Fig.~\ref{fig5} the dependence on the strain of the various contributions
to the energy coming from the different terms in the Hamiltonian.
It is quite evident that the various contributions to the total energy do NOT reach the same steady state, and
the result shown in Fig. \ref{fig4} is the consequence of an interesting and subtle cancellation that needs to be explained.
Currently we have no explanation to this observation. To be more confident in the correctness of the observation
we changed the parameter $\eta$ in Eq.~\ref{fene} and repeated the measurements; the observation remains invariant.

\subsection{Changes in geometry of polymers with applied strain}
\begin{figure}
 \includegraphics[scale=0.35]{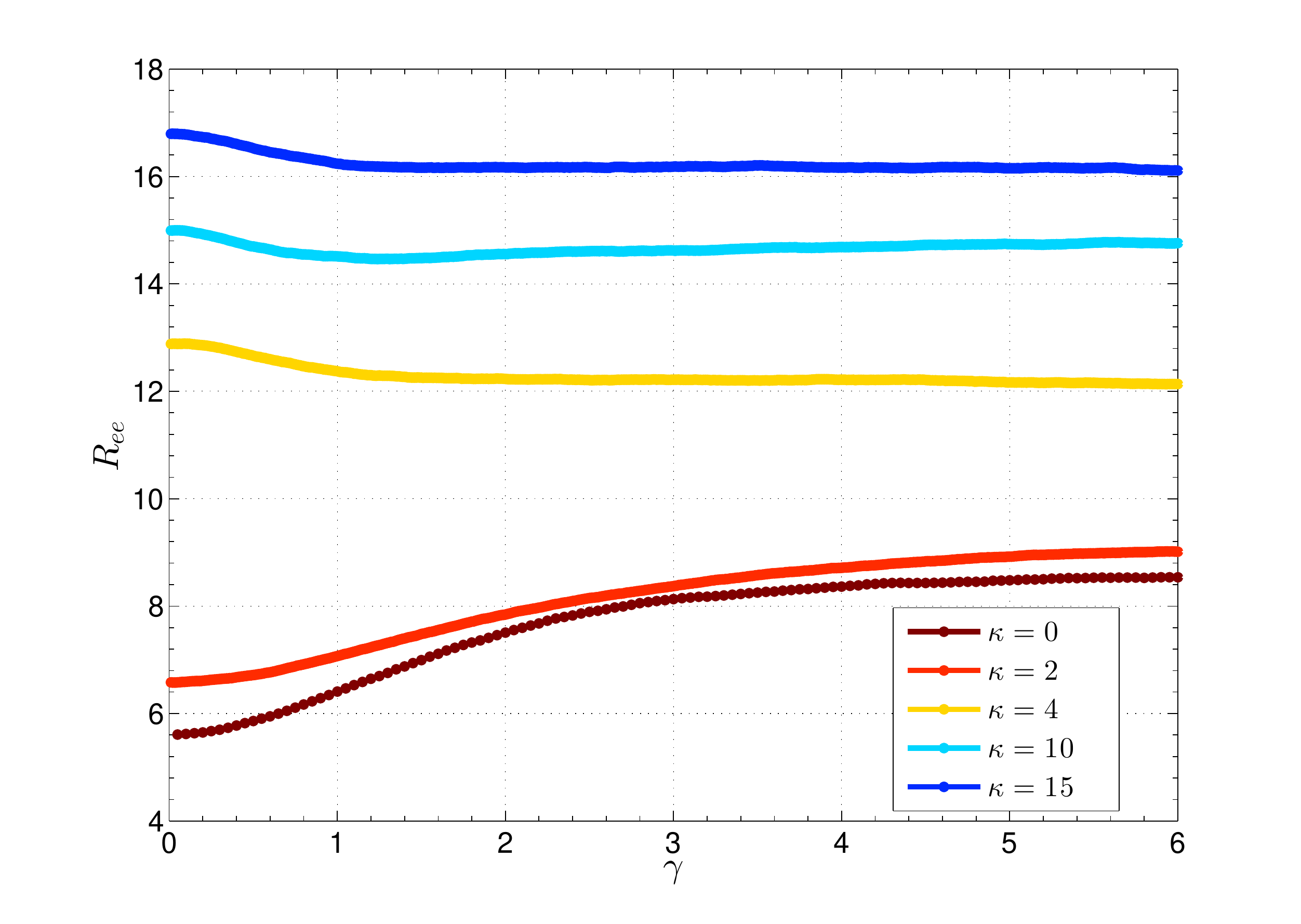}
 \caption{Variation of end to end length $R_{\rm ee}$ of the stiff polymers with the applied strain $\gamma$ as a function of $\kappa$ .}
 \label{fig6}
\end{figure}

\begin{figure}
 \includegraphics[scale=0.24]{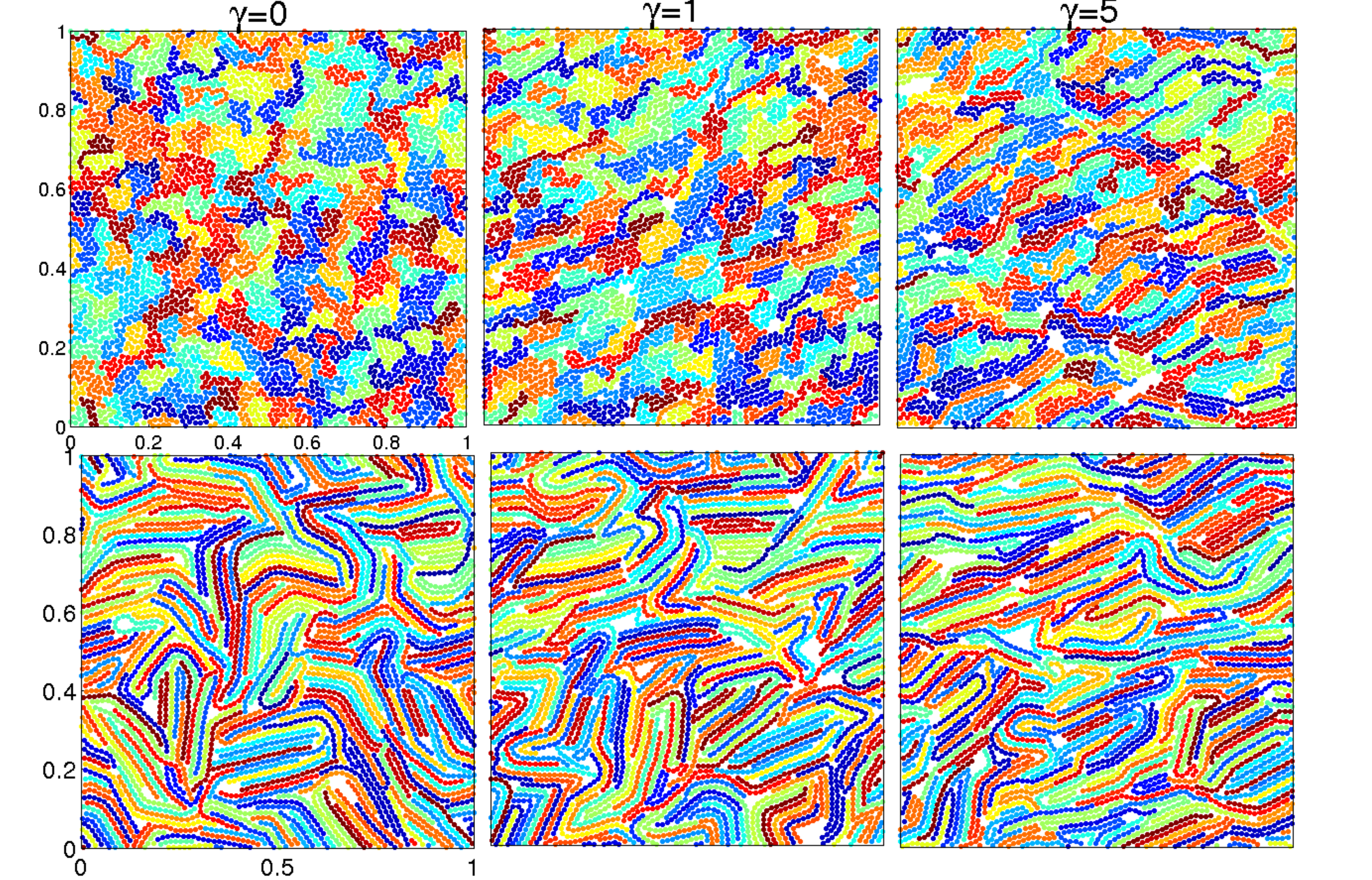}
 \caption{Snapshot of the polymer in the box for $\kappa=0$ (upper panel) and $\kappa=10$ (lower panel). From left to right  (a) $\gamma=0$, (b)$\gamma=1$, and (c) $\gamma=5$. For $\kappa=0$ the polymers are coiled for zero-strain and they stretch on average upon increasing of the strain. The opposite occurs for $\kappa=10$.}
 \label{fig7}
\end{figure}

In addition to the energy and the stress in the system, the oligomeric glass presents also interesting responses
to external strains in the resulting geometry of the chains. Of course,
the configuration of the oligomers in the glass depends on the stiffness of the chains. As the chains become stiffer the oligomer chain is easier to bend (since in our convention the straight chain is the minimal energy state). In order to characterize the configuration of the oligomer chains we compute the end-to-end length $R_{ee}$ of the chain and follow how it changes with the applied strain. Fig.~\ref{fig6} shows the variation of $R_{ee}$ for the stiff case as the applied strain is increased. We see that the tendency is different for small and large value of $\kappa$. For small $\kappa$ the
chains start from a coiled state, with $R_{ee}$ being of the order of $\sqrt{n}$. Then the action of the strain tends
to straighten the chains to increase $R_{ee}$ until a $\kappa$-dependent steady state.
On the other hand for large $\kappa$ one starts with almost straight chains, such the $R_{ee}$ is of the order of $n$;
straining now leads to bending, increasing the energy of the system, reaching again a $\kappa$-dependent steady state.
The process described can be seen directly in snapshots of the system under strain. This is shown for $\kappa=0$ and
$\kappa=10$ in Fig.~\ref{fig7}

\begin{figure}
 \includegraphics[scale=0.7]{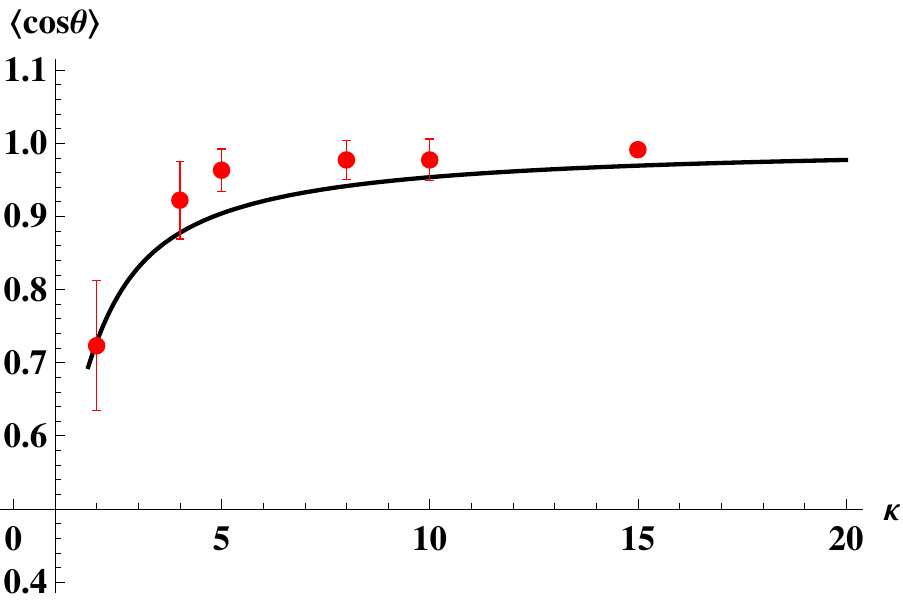}
 \includegraphics[scale=0.7]{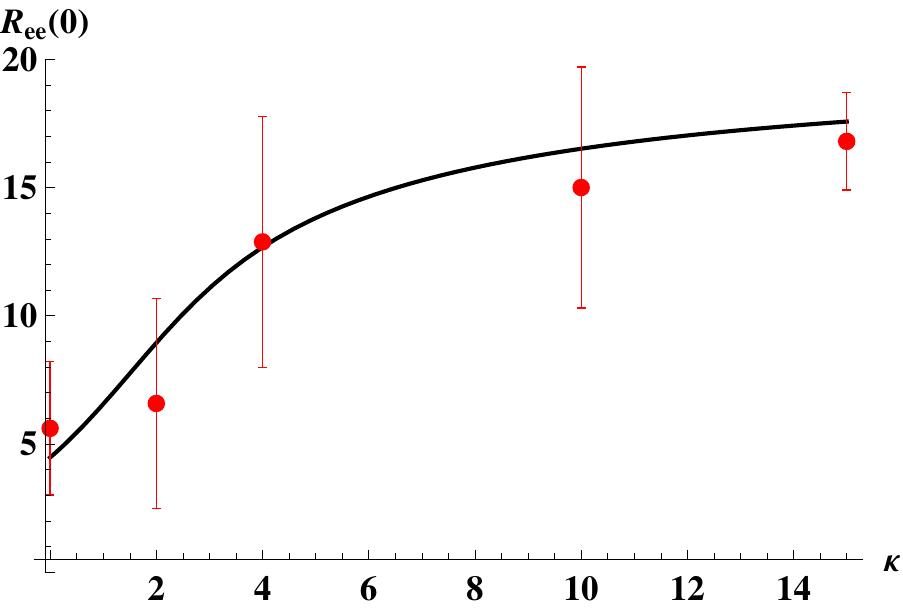}
 \includegraphics[scale=0.7]{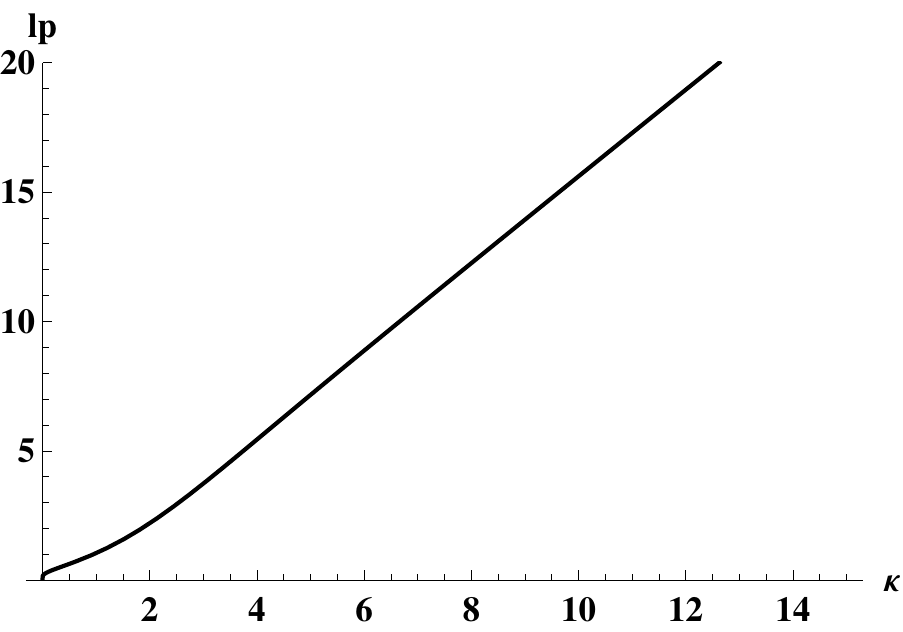}
 \caption{The average angular measure $\langle \cos \theta \rangle$ (upper panel), the average equilibrium end-to-end length $R_{ee}(0)$ (middle panel) and the average persistence length ($\ell_p$) (lower panel) as a function of $\kappa$ at zero strain ($\gamma=0$). The red dots with error bars are the data obtained from numerical simulations and black curves are the theoretical estimates obtained using the theory discussed in Sec.\ref{avcos}. }
 \label{fig8}
\end{figure}
\subsection{Theoretical remarks on the end-to-end distance and the average angular distribution}
\label{avcos}

The first observation that needs to be explained is the end-to-end distance in equilibrium (at $\gamma=0$). It turns out
that just using the angular potential is sufficient to give a good estimate of this distance. The reason is that because the
oligomers are fairly stiff, the Lennard Jones term does not lead to strong short-range particle-particle repulsion, while
the  main effect of the FENE term is simply to re-normalize the individual inter-bond distances. Thus using the angular potential
we can simply calculate the average angle of the oligomer chain which given by:
\begin{equation}\label{costh}
 \langle \cos\theta\rangle=\frac{\int^{\pi}_0 d\theta \cos\theta \exp\left(-\frac{U_{Angle}}{T}\right)}{\int^{\pi}_0 d\theta \exp\left(-\frac{U_{Angle}}{T}\right)} \ .
\end{equation}
Here the temperature $T$ should be taken to be of the order of the fluid melt from which the glass was quenched. Below we take
$T=1$. This integral can be performed exactly, and its value is $I_1(\kappa/T)/I_0(\kappa/T)$ where $I_1$ and $I_0$ are the
modified Bessel function of order 1 and 0 respectively. In the upper panel of Fig.~\ref{fig8} we compare the theoretical evaluation of $\langle \cos \theta\rangle$ to its numerically computed counterpart, and conclude the comparison is good.

Using the average angle we can write the average value $\langle R_{ee} \rangle$
\cite{AEGKP03} as:
\begin{equation}
\langle R_{ee}^2\rangle = n \left(\frac{1+\langle \cos \theta\rangle}{1-\langle \cos \theta\rangle}-\frac{1}{n}\frac{2\langle \cos \theta\rangle (1-\langle \cos \theta\rangle^n)}{(1-\langle \cos \theta\rangle)^2}\right)
\end{equation}
Taking the square-root of this expression we plot it
in the middle panel of Fig.~\ref{fig8} and compare it with the numerically
calculated value of $R_{ee}$ at $\gamma=0$, averaged over 40 different initial conditions. The agreement is quite acceptable.

Finally, it is advantageous to define `persistence length' $\ell_p$ using the relationship
\begin{equation}
\langle \cos \theta\rangle \equiv \exp(-1/\ell_p) \ .
\label{deflp}
\end{equation}
The resulting $\ell_p$ for $\gamma=0$ is shown in the lower panel of Fig.\ref{fig8}. We note that the persistence length
becomes of the order of $R_{ee}$ when the latter is about 10.

Returning to Fig.~\ref{fig6} one notes three interesting features:
\begin{enumerate}
\item For small values of $\kappa$ the end-to-end distance rises with increasing strain.
\item For large values of $\kappa$ the end-to-end distance decreases with increasing strain.
\item In either case the end-to-end distance attains a $\kappa$-dependent asymptotic value for large $\gamma$ which is
nevertheless {\em not} the fully stretched state.
\end{enumerate}

For $\kappa$ small there is a simple estimate of the asymptotic value of $R_{ee}$ which involves a balance between the force
due to straining which tends to stretch the oligomer and the entropic force which tends to keep the oligomer coiled. Estimating the
force due to stress as $\sigma R$ and the entropic force as $\frac{T}{R^2_{ee}} (R-R_{ee})$ \cite{Gennes}. Balancing the two
expressions we predict that
\begin{equation}
R_{ee} (\sigma)= \frac{R_{ee}(\sigma=0)}{1-\sigma R^2_{ee}(\sigma=0) /T} \ .
\label{Reesig}
\end{equation}
Indeed, the observed increase in the end-to-end distance at small values of $\kappa$ is in accordance
with this prediction. Of course for larger values of $\sigma$ the FENE terms need to be invoked to cure the apparent
divergence in Eq.~(\ref{Reesig}).

Once the persistence length is of the order of the initial value of $R_{ee}$ we can assume that the oligomers are
entirely stretched. Then the effect of the shear strain is opposite, in reducing the end-to-end distance. This stems
simply from the fact that any inclined stretched polymer will bend under the action of shear, since its two ends
move at different speeds. The reader can see this phenomenon occurring in the lower panel of Fig. \ref{fig7}.
Thus the effect of increasing $\gamma$ will initially decrease $R_{ee}$ as is observed
in Fig.~\ref{fig6}.

In both cases the estimate of the asymptotic value of $R_{ee}$ is not easy, and we leave
it for future research.
\section{Theory of plastic events}
\label{plastic}

The stability of amorphous solids is determined by the Hessian matrix which is made of second derivatives of the
Hamiltonian with respect to all the degrees of freedom. This matrix is always symmetric and real and therefore diagonalizable. As long as all the eigenvalues are positive, the system is mechanically stable. Plastic instabilities are characterized by an eigenvalue going to zero signaling the loss of mechanical stability.

\subsection{Calculation of the Hessian matrix}
To calculate the Hessian
matrix for the oligomeric glass we recognize the three contributions to the potential energy $\phi_{\rm LJ}$, $\chi$ and $\psi$. These contributions result in three sub matrices that need to be summed up to yield the full Hessian. We denote the sub matrices as $\mathcal H^{\rm LJ}$, $\mathcal H^{\rm FENE}$ and $\mathcal H^{\rm Angle}$:
\begin{equation}
\mathcal H=\mathcal H^{\rm LJ}+\mathcal H^{\rm FENE}+\mathcal H^{\rm Angle}
\end{equation}
We begin with $\C H^{\rm LJ}$:
\begin{eqnarray}\label{hessLJ}
\mathcal H^{\rm LJ}(i,j;\alpha,\beta)&=&\frac{\partial^2 \phi^{ij}_{\rm LJ}}{\partial x^j_\beta \partial x^i_\alpha}
=\frac{\partial^2 \phi_{\rm LJ}^{ij}}{\partial (r^{ij})^2}\frac{\partial r^{ij}}{\partial x^j_\beta}\frac{\partial r^{ij}}{\partial x^i_\alpha} \nonumber \\
&+&\frac{\partial \phi_{\rm LJ}^{ij}}{\partial r^{ij}}\frac{\partial^2 r^{ij}}{\partial x^j_\beta \partial x^i_\alpha},
\end{eqnarray}
and for $i=j$ it is:
\begin{equation}
\mathcal H^{\rm LJ}(i,i;\alpha,\beta)=\sum\limits_{\rm j\neq i}-\mathcal H^{\rm LJ}(i,j;\alpha,\beta).
\end{equation}
Note that unless otherwise stated latin letters (e.g., i, j, etc.) will be used for the particle's coordinate and greek letters (e.g., $\alpha,\beta$, etc.) will be used to the denote the displacement coordinate of the particles. In order to compute the terms used in the above equation (Eq.~\ref{hessLJ}) explicitly, we take the advantage of the identities:
\begin{equation}
\frac {\partial r^{m\ell}}{\partial x_\alpha^i}=\frac{r^{m\ell}_\alpha}{r^{m\ell}}(\delta^{\ell i}-\delta^{mi}),
\end{equation}
and
\begin{equation}
\frac{\partial^{2}r^{m\ell}}{\partial x_{\beta }^{j} \partial x_{\alpha }^{i}}=
\left( {\frac{\delta_{\alpha \beta } }{r^{m\ell}}-\frac{r_{\alpha }^{m\ell} r_{\beta }^{m\ell} }{(r^{m\ell})^{3}}} \right)\left( {\delta^{\ell j}-\delta^{mj}}\right)\left( {\delta^{\ell i}-\delta^{mi}} \right).
\end{equation}

Now we consider the terms in which the bond of a polymer connecting particles $k$ and $\ell$ contributes to the $\mathcal H^{\rm FENE}$. These terms are written as:
\begin{equation}
\mathcal H^{\rm FENE}(k,l;\alpha, \beta)=
\begin{pmatrix}
\frac{\partial^2 \chi^{k\ell}}{\partial x^k_\alpha \partial x^k_\beta} & \frac{\partial^2 \chi^{k\ell}}{\partial x^k_\alpha \partial x^\ell_\beta} \\
\frac{\partial^2 \chi^{k\ell}}{\partial x^\ell_\alpha \partial x^k_\beta} & \frac{\partial^2 \chi^{k\ell}}{\partial x^\ell_\alpha \partial x^\ell_\beta}
\end{pmatrix},
\end{equation}
where $\alpha$ and $\beta$ stand for all coordinates and thus the dimensions of $A$ are $2d\times 2d$. The four entries of matrix $A$ are not necessarily adjacent in $\mathcal H^{\rm FENE}$; depending on the values of $k$ and $\ell$ they are positioned at $(dk,dk)$, $(dk,d\ell)$, $(d\ell,dk)$ and $(d\ell,d\ell)$ respectively where $d$ is the dimensionality of the system.

Further we consider the terms related to the $i^{\rm th}$ valence angle within a polymer chain. This angle is defined by three successive particles with indices $k$,$\ell$ and $m$. The contribution of these terms to the Hessian $\mathcal H^{\rm Angle}$ is expressed as:
\begin{equation}
\mathcal H^{\rm Angle}=
\begin{pmatrix}
\frac{\partial^2 \psi^{i}}{\partial x^{k}_\alpha \partial x^{k}_\beta} & \frac{\partial^2 \psi^{i}}{\partial x^{k}_\alpha \partial x^{\ell}_\beta} & \frac{\partial^2 \psi^{i}}{\partial x^{k}_\alpha \partial x^{m}_\beta} \\
\frac{\partial^2 \psi^{i}}{\partial x^{\ell}_\alpha \partial x^{k}_\beta} & \frac{\partial^2 \psi^{i}}{\partial x^{\ell}_\alpha \partial x^{\ell}_\beta} & \frac{\partial^2 \psi^{i}}{\partial x^{\ell}_\alpha \partial x^{m}_\beta} \\
\frac{\partial^2 \psi^{i}}{\partial x^{m}_\alpha \partial x^{k}_\beta} & \frac{\partial^2 \psi^{i}}{\partial x^{m}_\alpha \partial x^{\ell}_\beta} & \frac{\partial^2 \psi^{i}}{\partial x^{m}_\alpha \partial x^{m}_\beta}
\end{pmatrix},
\end{equation}
where $\alpha$ and $\beta$ stand for all coordinates and thus the dimensions of $\mathcal H_{\rm Angle}$ are $3d\times 3d$. The four entries of $\mathcal H_{\rm Angle}$ are not necessarily adjacent, depending on the values of $k$, $\ell$ and $m$ they are positioned at $(dk,dk)$, $(dk,d\ell)$, $(dk,dm)$, $(d\ell,dk)$, $(d\ell,d\ell)$, $(d\ell,dm)$, $(dm,dk)$, $(dm,d\ell)$ and $(dm,dm)$, respectively, where $d$ is dimensionality of the system.

Formally the angular contribution to the Hessian can be expressed as:

\begin{eqnarray}\label{hessang}
\mathcal H_{\rm Angle}(i,j;\alpha,\beta)&=&\frac{\partial^{2}\psi^{\ell}}{\partial x_{\beta }^{j} \partial x_{\alpha}^{i} }
=\frac{\partial^{2}\psi^{\ell}}{(\partial \cos \varphi
^{\ell})^{2}}\frac{\partial \cos \varphi^{\ell}}{\partial x_{\alpha}^{i}
}\frac{\partial \cos \varphi^{\ell}}{\partial x_{\beta }^{j} }\nonumber \\
&+&\frac{\partial
\psi^{\ell}}{\partial \cos \varphi^{\ell}}\frac{\partial^{2}\cos \varphi
^{\ell}}{ \partial x_{\beta }^{j} \partial x_{\alpha }^{i}},
\end{eqnarray}

where the cosine of the valence angle is defined as:
\[
\cos \varphi^{l}=-\frac{r_{\gamma }^{l-1,l} r_{\gamma }^{l,l+1}
}{r^{l-1,l} r^{l,l+1} },
\]
with $r_{\gamma }^{kl} =r_{\gamma }^{l} -r_{\gamma}^{k}$. Now inserting the formula for cosine in the Eq.~\ref{hessang}, one obtains:
\begin{widetext}
\[
\begin{array}{l}
 \frac{\partial \cos \varphi^{l}}{\partial x_{\alpha }^{m} }=-\left[
{\frac{r_{\gamma }^{l,l+1} }{r^{l,l+1} }\frac{\partial }{\partial
x_{\alpha }^{m} }\left( {\frac{r_{\gamma }^{l-1,l} }{r^{l-1,l} }}
\right)+\frac{r_{\gamma }^{l-1,l} }{r^{l-1,l} }\frac{\partial }{\partial
x_{\alpha }^{m} }\left( {\frac{r_{\gamma }^{l,l+1} }{r^{l,l+1} }}
\right)} \right] \\
 \frac{\partial^{2}\cos \varphi^{l}}{\partial x_{\alpha }^{m} \partial
x_{\beta }^{q} }=-\left[ {\begin{array}{l}
 \frac{r_{\gamma }^{l,l+1} }{r^{l,l+1} }\frac{\partial^{2}}{\partial
x_{\alpha }^{m} \partial x_{\beta }^{q} }\left( {\frac{r_{\gamma }^{l-1,l}
}{r^{l-1,l} }} \right)+\frac{r_{\gamma }^{l-1,l} }{r^{l-1,l}
}\frac{\partial^{2}}{\partial x_{\alpha }^{m} \partial x_{\beta }^{q}
}\left( {\frac{r_{\gamma }^{l,l+1} }{r^{l,l+1} }} \right)+ \\
 +\frac{\partial }{\partial x_{\alpha }^{m} }\left( {\frac{r_{\gamma
}^{l-1,l} }{r^{l-1,l} }} \right)\frac{\partial }{\partial x_{\beta }^{q}
}\left( {\frac{r_{\gamma }^{l,l+1} }{r^{l,l+1} }} \right)+\frac{\partial
}{\partial x_{\beta }^{q} }\left( {\frac{r_{\gamma }^{l-1,l} }{r^{l-1,l}
}} \right)\frac{\partial }{\partial x_{\alpha }^{m} }\left( {\frac{r_{\gamma
}^{l,l+1} }{r^{l,l+1} }} \right) \\
 \end{array}} \right] \\
 \end{array}
\]

The other auxiliary expressions are given by:

\[
\begin{array}{l}
 \frac{\partial }{\partial x_{\alpha }^{m} }\left( {\frac{r_{\gamma }^{kl}
}{r^{kl}}} \right)=\left( {\frac{\delta_{\alpha \gamma }
}{r^{kl}}-\frac{r_{\alpha }^{kl} r_{\gamma }^{kl} }{(r^{kl})^{3}}}
\right)\left( {\delta^{lm}-\delta^{km}} \right) \\
 \frac{\partial^{2}}{\partial x_{\alpha }^{m} \partial x_{\beta }^{q}
}\left( {\frac{r_{\gamma }^{kl} }{r^{kl}}} \right)=\left( {\frac{3r_{\alpha
}^{kl} r_{\beta }^{kl} r_{\gamma }^{kl} }{(r^{kl})^{5}}-\frac{\delta
_{\alpha \beta } r_{\gamma }^{kl} +\delta_{\alpha \gamma } r_{\beta }^{kl}
+\delta_{\beta \gamma } r_{\alpha }^{kl} }{(r^{kl})^{3}}} \right)\left(
{\delta^{lm}-\delta^{km}} \right)\left( {\delta^{lq}-\delta^{kq}}
\right) \\
 \end{array}
\].

Combining the above expressions together we have:

\[
\begin{array}{l}
 \frac{\partial \cos \varphi^{l}}{\partial x_{\alpha }^{m} }=-\left[
{\frac{r_{\gamma }^{l,l+1} }{r^{l,l+1} }\frac{\partial }{\partial
x_{\alpha }^{m} }\left( {\frac{r_{\gamma }^{l-1,l} }{r^{l-1,l} }}
\right)+\frac{r_{\gamma }^{l-1,l} }{r^{l-1,l} }\frac{\partial }{\partial
x_{\alpha }^{m} }\left( {\frac{r_{\gamma }^{l,l+1} }{r^{l,l+1} }}
\right)} \right]= \\
 =-\left[ {\left( {\delta^{l,m}-\delta^{l-1,m}} \right)\left(
{\frac{r_{\alpha }^{l,l+1} }{r^{l,l+1} r^{l-1,l} }-\frac{r_{\alpha
}^{l-1,l} r_{\gamma }^{l-1,l} r_{\gamma }^{l,l+1} }{r^{l,l+1}
(r^{l-1,l} )^{3}}} \right)+\left( {\delta^{l+1,m}-\delta^{l,m}}
\right)\left( {\frac{r_{\alpha }^{l-1,l} }{r^{l,l+1} r^{l-1,l}
}-\frac{r_{\alpha }^{l,l+1} r_{\gamma }^{l-1,l} r_{\gamma }^{l,l+1}
}{r^{l-1,l} (r^{l,l+1} )^{3}}} \right)} \right] \\
 \end{array},
\]

$\begin{array}{l}
 \frac{\partial^{2}\cos \varphi^{l}}{\partial x_{\alpha }^{m} \partial
x_{\beta }^{q} }=-\left[ {\begin{array}{l}
 \frac{r_{\gamma }^{l,l+1} }{r^{l,l+1} }\frac{\partial^{2}}{\partial
x_{\alpha }^{m} \partial x_{\beta }^{q} }\left( {\frac{r_{\gamma }^{l-1,l}
}{r^{l-1,l} }} \right)+\frac{r_{\gamma }^{l-1,l} }{r^{l-1,l}
}\frac{\partial^{2}}{\partial x_{\alpha }^{m} \partial x_{\beta }^{q}
}\left( {\frac{r_{\gamma }^{l,l+1} }{r^{l,l+1} }} \right)+ \\
 +\frac{\partial }{\partial x_{\alpha }^{m} }\left( {\frac{r_{\gamma
}^{l-1,l} }{r^{l-1,l} }} \right)\frac{\partial }{\partial x_{\beta }^{q}
}\left( {\frac{r_{\gamma }^{l,l+1} }{r^{l,l+1} }} \right)+\frac{\partial
}{\partial x_{\beta }^{q} }\left( {\frac{r_{\gamma }^{l-1,l} }{r^{l-1,l}
}} \right)\frac{\partial }{\partial x_{\alpha }^{m} }\left( {\frac{r_{\gamma
}^{l,l+1} }{r^{l,l+1} }} \right) \\
 \end{array}} \right]= \\
 =-\left[ {\begin{array}{l}
 \left( {\frac{3r_{\alpha }^{l-1,l} r_{\beta }^{l-1,1} r_{\gamma }^{l-1,l}
r_{\gamma }^{l,l+1} }{r^{l,l+1}(r^{l-1,l})^{5}}-\frac{\left[ {\delta
_{\alpha \beta } r_{\gamma }^{l-1,l} r_{\gamma }^{l,l+1} +r_{\alpha
}^{l,l+1} r_{\beta }^{l-1,1} +r_{\alpha }^{l-1,1} r_{\beta }^{l,l+1} }
\right]}{r^{l,l+1}(r^{l-1,l})^{3}}} \right)\left( {\delta^{lm}-\delta
^{l-1,m}} \right)\left( {\delta^{lq}-\delta^{l-1,q}} \right)+ \\
 +\left( {\frac{3r_{\alpha }^{l,l+1} r_{\beta }^{l,1+1} r_{\gamma }^{l-1,l}
r_{\gamma }^{l,l+1} }{r^{l-1,l}(r^{l,l+1})^{5}}-\frac{\left[ {\delta
_{\alpha \beta } r_{\gamma }^{l-1,l} r_{\gamma }^{l,l+1} +r_{\alpha
}^{l,l+1} r_{\beta }^{l-1,1} +r_{\alpha }^{l-1,1} r_{\beta }^{l,l+1} }
\right]}{r^{l-1,l}(r^{l,l+1})^{3}}} \right)\left( {\delta^{l+1,m}-\delta
^{l,m}} \right)\left( {\delta^{l+1,q}-\delta^{l,q}} \right)+ \\
 +\left( {\frac{\delta_{\alpha \beta }
}{r^{l-1,l}r^{l,l+1}}-\frac{r_{\alpha }^{l-1,l} r_{\beta }^{l-1,1}
}{r^{l,l+1}(r^{l-1,l})^{3}}-\frac{r_{\alpha }^{l,l+1} r_{\beta }^{l,1+1}
}{r^{l-1,l}(r^{l,l+1})^{3}}+\frac{r_{\alpha }^{l-1,l} r_{\beta }^{l,1+1}
r_{\gamma }^{l-1,l} r_{\gamma }^{l,l+1} }{(r^{l-1,l})^{3}(r^{l,l+1})^{3}}}
\right)\left( {\delta^{l,m}-\delta^{l-1,m}} \right)\left( {\delta
^{l+1,q}-\delta^{l,q}} \right)+ \\
 +\left( {\frac{\delta_{\alpha \beta }
}{r^{l-1,l}r^{l,l+1}}-\frac{r_{\alpha }^{l-1,l} r_{\beta }^{l-1,1}
}{r^{l,l+1}(r^{l-1,l})^{3}}-\frac{r_{\alpha }^{l,l+1} r_{\beta }^{l,1+1}
}{r^{l-1,l}(r^{l,l+1})^{3}}+\frac{r_{\alpha }^{l,l+1} r_{\beta }^{l-1,1}
r_{\gamma }^{l-1,l} r_{\gamma }^{l,l+1} }{(r^{l-1,l})^{3}(r^{l,l+1})^{3}}}
\right)\left( {\delta^{l+1,m}-\delta^{l,m}} \right)\left( {\delta
^{l,q}-\delta^{l-1,q}} \right) \\
 \end{array}} \right] \\
 \end{array}.
$
\end{widetext}
\subsection{Elementary plastic events}

\begin{figure}
 \includegraphics[scale=0.41]{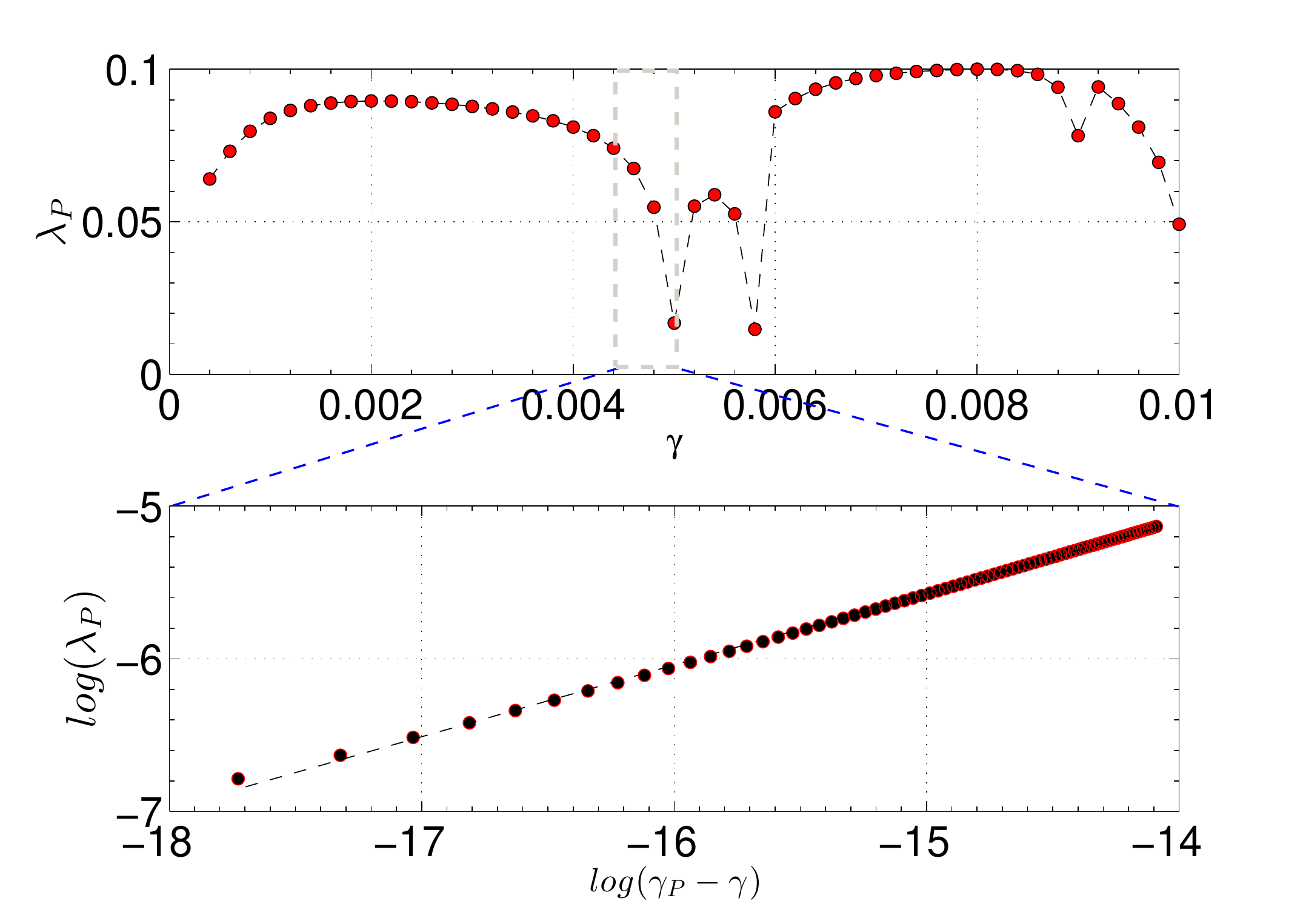}
 \caption{The variation of the smallest eigenvalue $\lambda_P$ as $\gamma$ is increased, In the upper panel
 we see the eigenvalue dips to zero, then recovers after the instability is over, and again dips to zero at the
 next instability. In the lower panel we choose to blow up the region of the first instability to demonstrate the
 approach of the eigenvalue to zero with a square-root singularity Eq. (\ref{sr}).}
 \label{fig11}
 \end{figure}

\begin{figure}
 \includegraphics[scale=0.5]{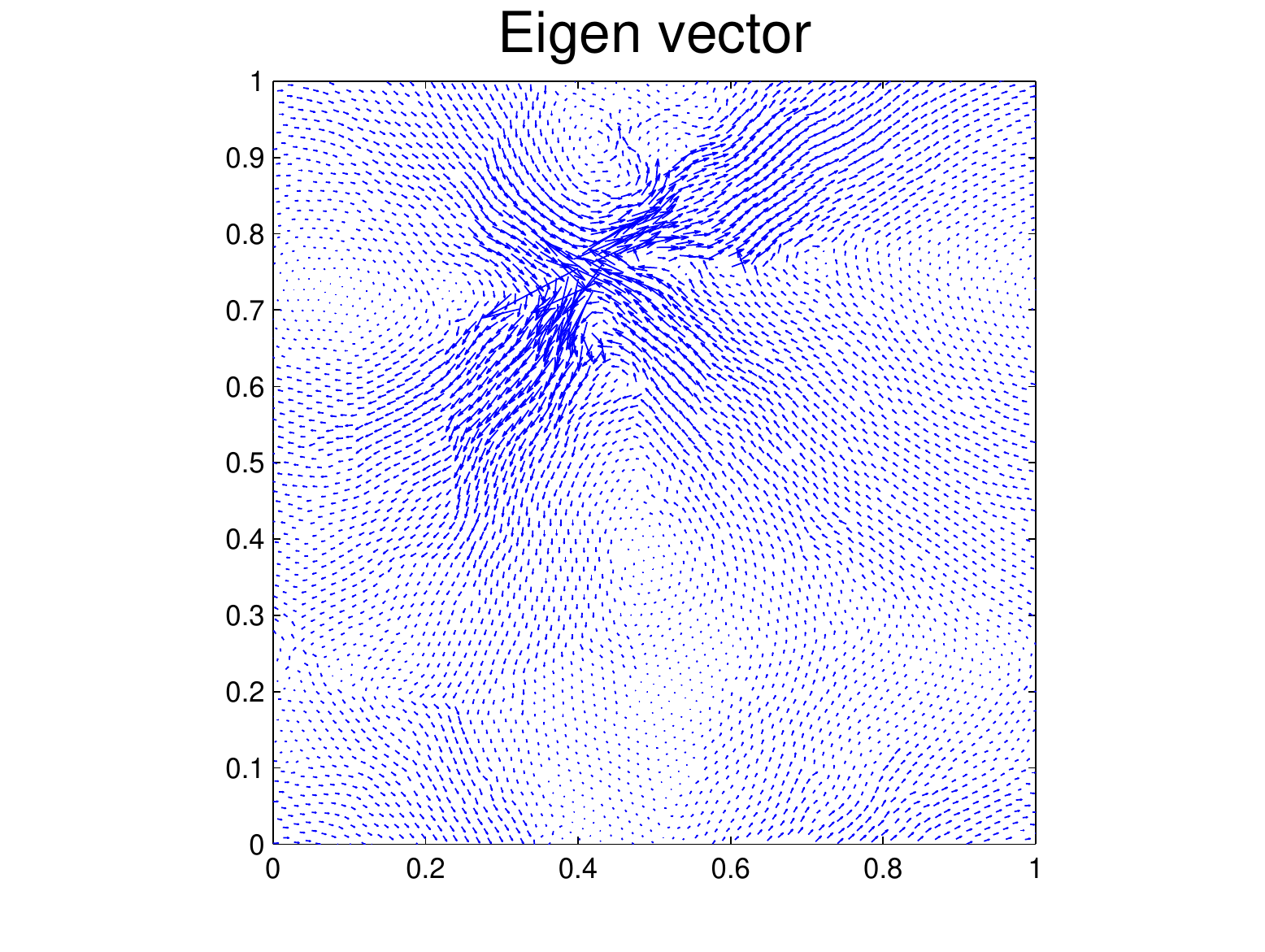}
 \includegraphics[scale=0.5]{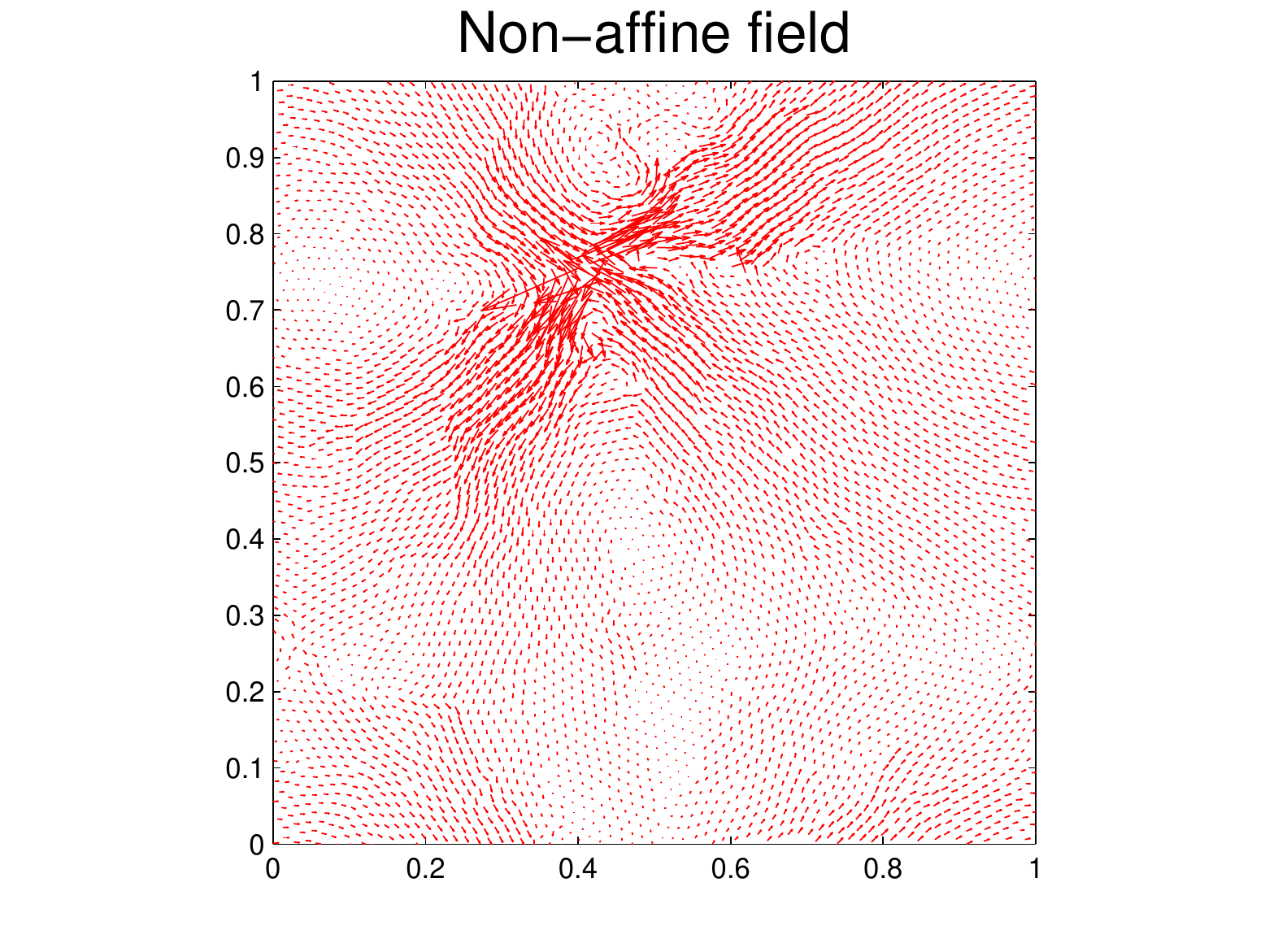}
 \caption{The eigenvector $\Psi_P$ and the non-affine displacement field associated with the first
 plastic instability as $\lambda\to 0$.}
 \label{fig12}
 \end{figure}

\begin{figure}
 \includegraphics[scale=0.50]{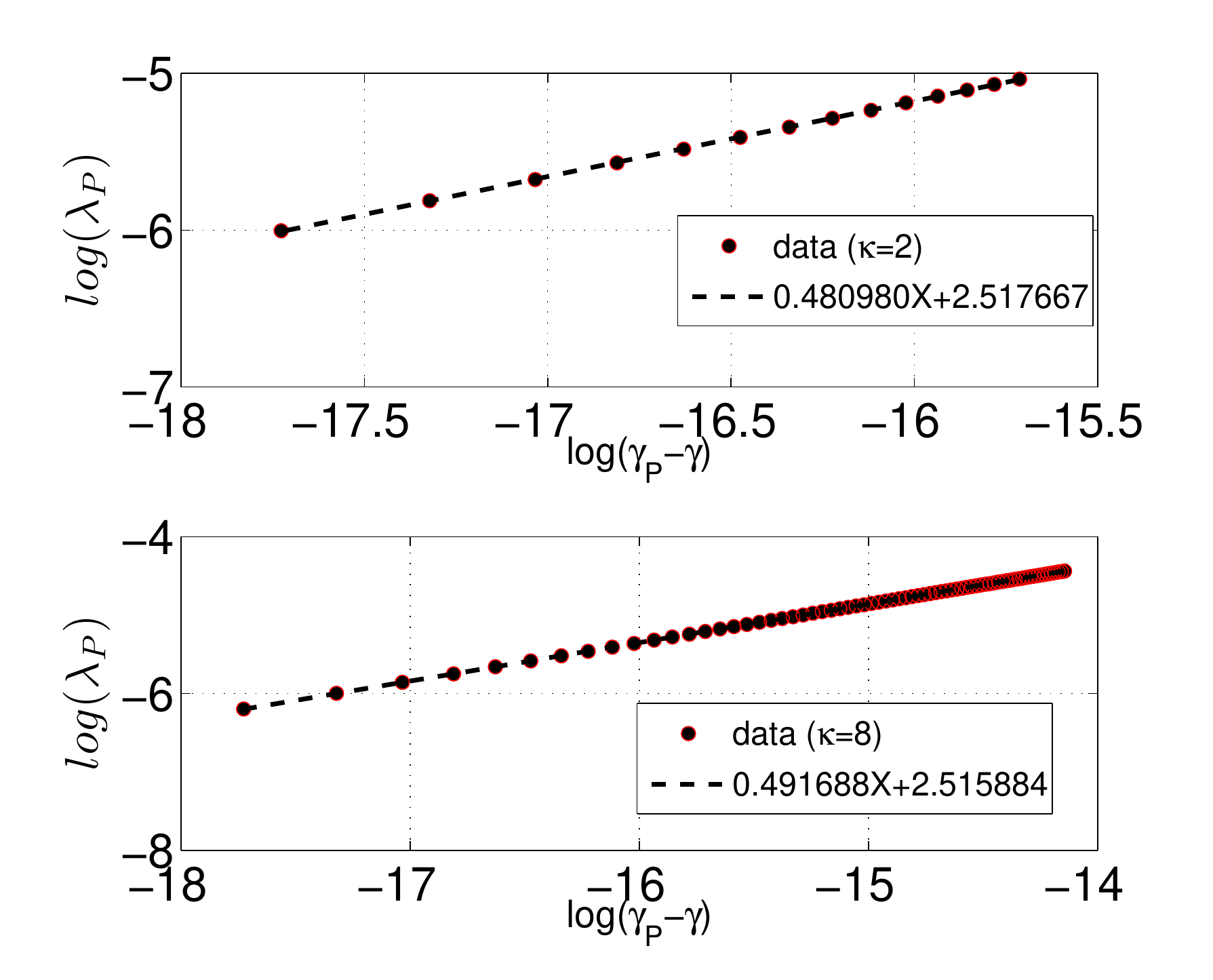}
 \caption{Log-log  plot of the eigen value of the plastic mode $\lambda_P$ Vs $\gamma_P-\gamma$ for $\kappa=2 $ (top panel) and for $\kappa$=8 (bottom panel) near the first elementary plastic event for semi-flexible ($\alpha=2$) polymer. The exponent is approximately $0.5$.  }
 \label{fig13}
\end{figure}

Having calculated the Hessian matrix we can now examine the elementary plastic events that occur at small
values of $\gamma$. As said above, the mechanical stability is lost when an eigenvalue of the Hessian goes to
zero. This is occurring via a saddle-node bifurcation in which the minimum in which the system resides collides
with a saddle of the global energy surface. During a saddle node bifurcation the approach of the eigenvalue to
zero is generic, following a square-root singularity~\cite{LP09}
\begin{equation}
\lambda_p \sim \sqrt{\gamma_p-\gamma} \ , \label{sr}
\end{equation}
where $\lambda_P$ is the eigenvalue that reaches zero at $\gamma=\gamma_P$. An example of this square-root singularity
for a stiff oligomeric glass with $\kappa=2$ is shown in Fig. \ref{fig11}.
 As the instability is approached the non-affine response becomes closer to the eigenvector of the Hessian matrix
 that is associated with $\lambda_P$, denoted as $\Psi_P$. This phenomenon is demonstrated in Fig. \ref{fig12}.
It is important to stress that the square-root singularity is generic and characteristic to saddle-node bifurcations.
It should be therefore independent of the system parameters and even the nature of the system. In our case we demonstrate this universality by changing form stiff to semi-flexible and measuring the eigenvalue $\lambda_P$ for two
values of $\kappa$ as shown in Fig.~\ref{fig13}.

\section{Formation of shear bands}
\label{failure}

Oligomeric glasses, like simple binary glasses and the much more complex metallic glasses, exhibit, in addition to
localized plastic events also a second class of system spanning, shear localizing events. These events are precursors
to shear banding, and they need a finite amount of stress or strain to accumulate before they become possible. In previous analysis it was shown that shear localizing events occur when the strain exceeds a value $\gamma_{_{\rm Y}}$
which depends on the Poisson ratio of the material but is usually around 5-7\%~\cite{DHP12}. It appears that the present oligomeric
glasses are not much different in this respect. We begin to see shear localizing instabilities when $\gamma$ is of
the order of 10\% or less. The shear localization event is rather dramatic; even though we shear homogeneously with our
affine transformation the system chooses to respond by localizing all the shear over a small band of the size of the core of the Eshelby solution, and see Refs.~\cite{DHP12,DGMPS13} for details. It was shown in Refs.~\cite{DHP12,DGMPS13} that this solution minimizes the energy
compared to a random array of elementary plastic events.

The nature of the shear localizing events is similar to what had been seen previously: an eigenvalue of the Hessian matrix dips to zero, but now instead of a single quadrupolar structure a whole string of those, concatenated along a line in 2-dimensions~\cite{DHP12} or on a plane in 3-dimensions~\cite{DGMPS13}, appear simultaneously. They have a global connection now, with the outgoing direction of one quadrupolar structure connecting immediately to the incoming direction of the next quadrupole, thus arranging the displacement field to go in two different direction above and below the line (or plane). For pure shear the line (or plane) is in 45$^o$ to the principal stress axis. Other angles are possible for uniaxial loading~\cite{JGPS13}.

\begin{figure}
 \includegraphics[scale=0.5]{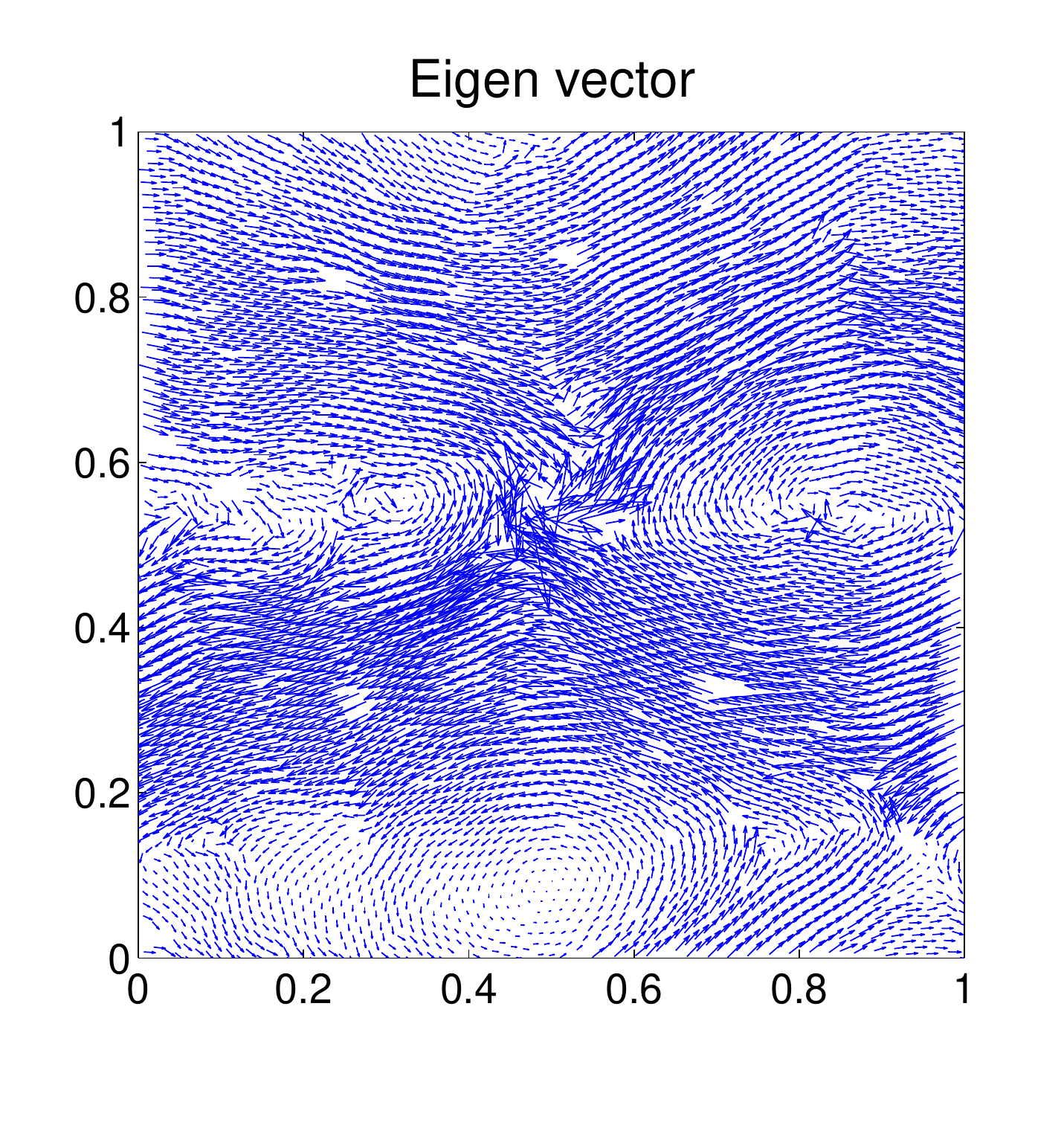}
 \includegraphics[scale=0.5]{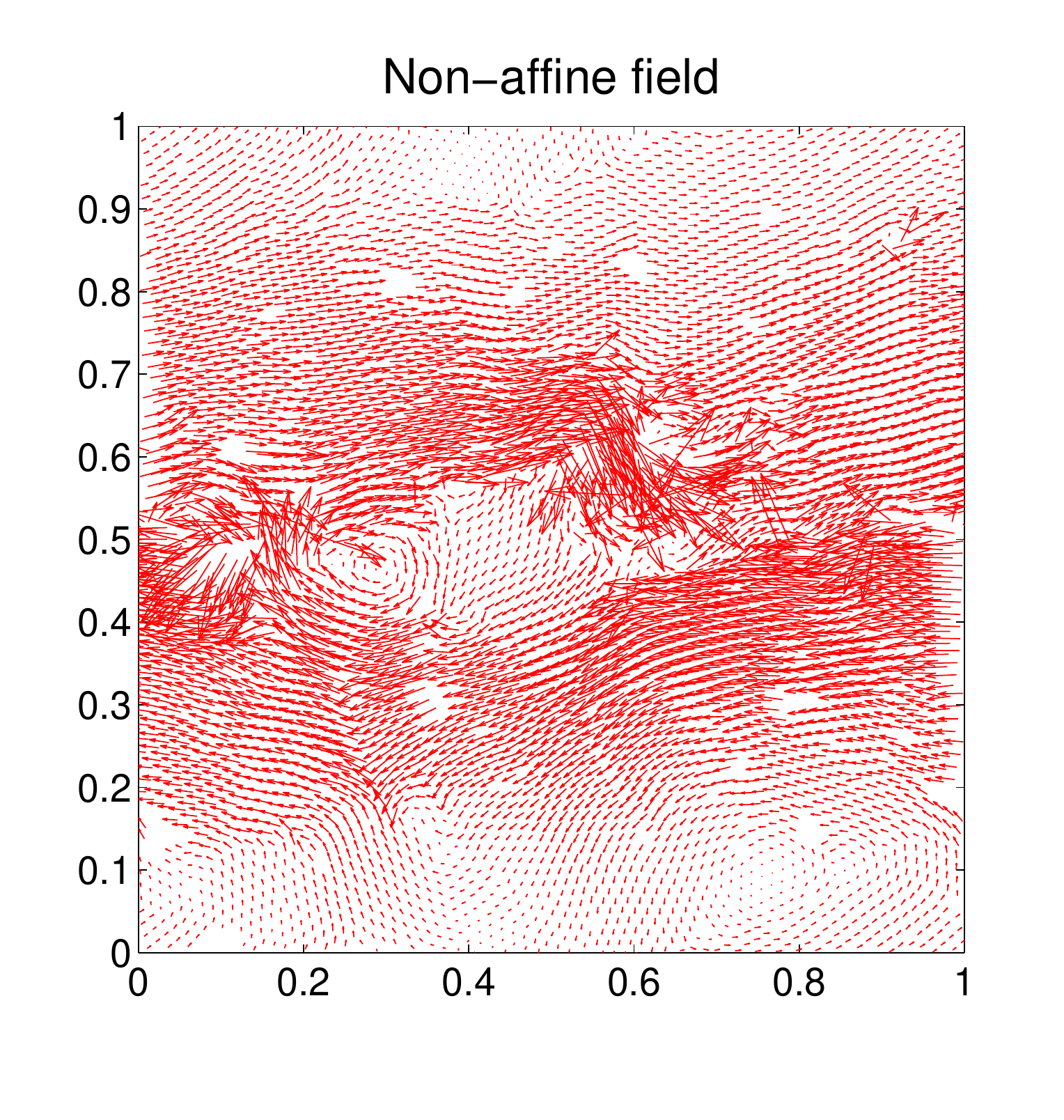}
 \caption{The eigenfunction $\Psi_P$ and the non-affine displacement field associated with a shear localizing
 plastic instability as $\lambda\to 0$. Note the global connection between the series of quadrupoles arranged along
 the line, such that the displacement field is pointing right above and left below the line. This IS the phenomenon of
 shear localization.}
 \label{fig14}
 \end{figure}

The best way to demonstrate the phenomenon is to display the eigenfunction or the displacement field associated with the event. In Fig. \ref{fig14} we show both, the eigenfunction in the upper panel and the directly simulated
 non-affine displacement field at the instability in the lower panel. Both images show how the shear is now
 concentrated over a narrow band, with the displacement field pointing to the ``right" above the band and to
 the ``left" below the band. In a stress control rather than a strain controlled experiment such an event
 would lead to macroscopic failure.

 In the next section we will present a theoretical formalism to compute the shear modulus for the oligomeric glasses.

 \section{Shear modulus $\mu$}
 \label{modulus}

The shear modulus that is a measure of  linear response of the material under the applied strain characterizes the mechanical behaviour of the system. Here  we provide the theory that relates the shear modulus to the microscopic variables like Hessian, non-affine displacements, etc.

We recall that for homogeneous shear strain the shear modulus is defined as the second derivatives of the potential energy with respect to the applied strain $\gamma$, i.e.,
\begin{equation}
 \mu=\frac{1}{V}\frac{d^2U(r_1,r_2,....,r_N;\gamma)}{d\gamma^2}.
 \label{mu-1}
\end{equation}
In this expression the second derivative contains two contributions: one coming from the affine part and another from the non-affine motion of the monomers. Thus we have~\cite{10KLP}
\begin{equation}\label{dGamm}
 \frac{d}{d\gamma}=\frac{\partial}{\partial \gamma}+\frac{\partial}{\partial \mathbf{u_i}}\cdot\frac{\partial \mathbf{u_i}}{\partial \gamma}\equiv \frac{\partial}{\partial \mathbf{r_i}}\cdot\frac{\partial \mathbf{u_i}}{\partial \gamma},
\end{equation}
where the second equality follows from the relation: $d\mathbf{r_i}=d\mathbf{u_i}$. Now the expression for shear modulus has the form:
\begin{equation}
 \mu=\frac{\partial^2 U}{\partial \gamma^2}+\frac{\partial u_i}{\partial \gamma}\frac{\partial U}{\partial r_i \partial \gamma}.
 \label{mu-2}
\end{equation}

Further we note that the affine step is followed by the non-affine step that returns the system to the equilibrium state. For the equilibrium state

\begin{equation}\label{forceEq}
 \frac{df_i}{d\gamma}\equiv-\frac{d}{d\gamma}\frac{\partial U}{\partial r_i}=0,
\end{equation}
where $f_i$ is the force on the $i^{\rm th}$ particle. As we use the Eq.~\ref{dGamm} in the above equation (Eq.~\ref{forceEq}) we obtain
\begin{equation}
 \frac{d\mathbf {u_i}}{d \gamma}= -H_{ij}^{-1}\cdot \Xi_j,
 \label{non-affine-vel}
\end{equation}
where $H_{ij}$ is the Hessian and $\Xi_j=\frac{\partial^2 U}{\partial\gamma \partial r_j}$ is the non-affine force. Now putting back Eq.(\ref{non-affine-vel}) into Eq.(\ref{mu-2}) we obtain the expression for shear modulus as

\begin{equation}
 \mu= \frac{1}{V}\frac{\partial^2 U(r_1,r_2,....r_n;\gamma)}{\partial\gamma^2}-\frac{1}{V}\sum_{i,j}\Xi_i\cdot H^{-1}_{ij}\cdot\Xi_{j}.
\end{equation}
The first term in the above expression represents contribution in the shear modulus  as a result of the affine displacement (also called as Born term), while  the second one is the contribution due to the non-affine responses. The Born term is computed analytically in Appendix \ref{Born}.
The so called ``non-affine force" $\B \Xi$ is calculated directly from the knowledge of the potential, see the Appendix, and then we solve the inverted equation (\ref{non-affine-vel}) $\B H \cdot \frac{d\B u}{d\gamma} = \B \Xi$ using
conjugate gradient minimization.  Having at hand the non-affine velocity $d\B u/d\gamma$ we can get the non-affine
contribution to the shear modulus using Eq. (\ref{mu-2}).

\begin{figure}
 \includegraphics[scale=0.5]{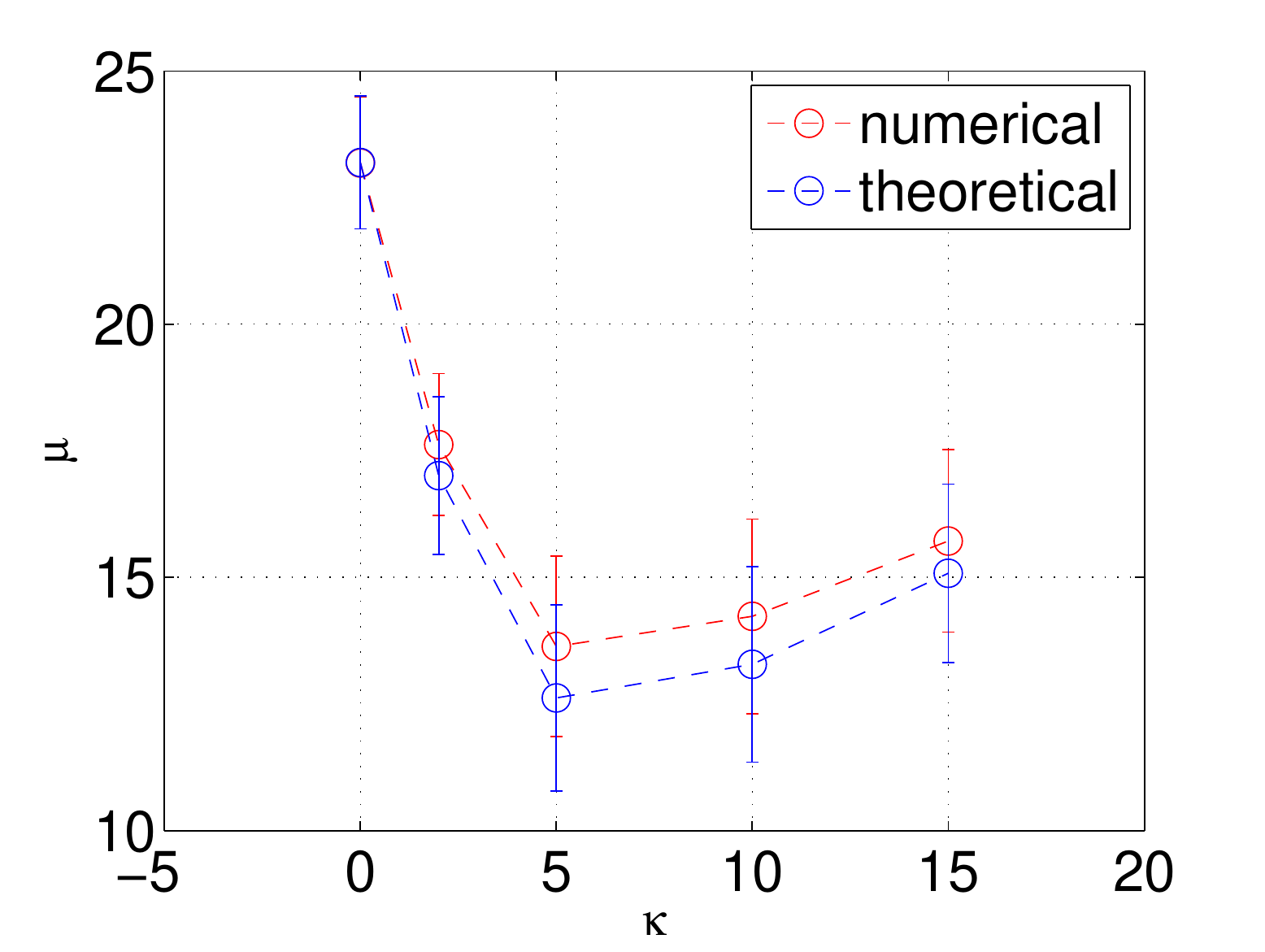}
\caption{
Comparison of the theoretically calculated and the numerically estimated shear modulus for various
values of $\kappa$. The relatively large error bars stem from the relative smallness of the system,
in which different realizations give a spread of values of the shear modulus. Nevertheless the agreement
between theory and simulations is quite satisfactory.}
\label{fig15}
\end{figure}

A comparison between the theoretically calculated shear modulus and the one estimated directly from the
stress vs. strain curves at very small $\gamma$ is provided in Fig. \ref{fig15}.

\section{Summary and concluding remarks}

In this paper we discussed the mechanics of oligomeric glasses, also known as waxes, with a special attention to the stress and
energy vs strain, the characteristics of the oligomeric chains and their changes under strain, the shear modulus, and the
plastic failure modes. We proposed a microscopic outlook which extends the available theory for simple binary glasses to this
much more complex oligomeric example. This resulted in an exact theory for the shear modulus, and a full understanding of the
plastic failure, both in the localized and the extended modes.

There are a few open problems that call for further theoretical and numerical considerations. The most relevant are:
\begin{enumerate}
\item A continuum theory of the stress vs strain and energy vs strain is lacking. To be realistic, this is a hard task,
and even for the simpler case of binary glasses such a theory is still under hard debate \cite{FL11,00FSC}. Understanding the energy budget
will be crucial in achieving progress along these lines.

\item A theory of the conformational changes of the oligomeric chain under strain is missing. We have provided above a theory
of the end-to-end distance for the case $\gamma=0$ but not for finite $\gamma$.

\item The extension of the approach to three dimensions is highly desirable. There one can expect interesting effects of
oligomer interpenetration, trapping and reptation, especially with longer oligomers and under higher strains.

\end{enumerate}

At least the last of these open issues is under active study in our laboratory, and we hope to present it in the near future.

\appendix

\section{Analytic computation of Born term}
\label{Born}

For small strain field the potential energy can be expressed as:

\begin{equation}
U=U_0+\frac{\partial U}{\partial \epsilon_{\alpha\beta}}\epsilon_{\alpha \beta}+\frac{1}{2}\frac{\partial^2 U}{\partial\epsilon_{\alpha\beta}\partial\epsilon_{\eta\nu}}\epsilon_{\alpha\beta}\epsilon_{\eta\nu}+\mathcal O (\epsilon^3).
\end{equation}

Also for the simple shear with affine transformation $h$ we have:

\begin{equation}
h^{T}h=
\begin{pmatrix}
1 & 0 \\
\gamma & 1
\end{pmatrix}
\cdot
\begin{pmatrix}
1 & \gamma\\
0 & 1
\end{pmatrix} =
\begin{pmatrix}
1 & \gamma \\
\gamma & 1+\gamma^2
\end{pmatrix} =
2\epsilon+I_2.
\end{equation}
 Thus the strain field $\epsilon$ can be written as:

\begin{equation}\epsilon=
\begin{pmatrix}
0 & \gamma/2 \\
\gamma/2 & \gamma^2/2
\end{pmatrix}
\end{equation}

The Born contribution to the shear modulus $\mu_B$ can be expressed as:

\begin{equation}
\mu_B\equiv \frac{d^2 U}{d \gamma^2}=\frac{\partial^2 U}{\partial \epsilon^2_{xy}}+\frac{\partial U}{\partial \epsilon_{yy}}
\end{equation}

At this stage we need to compute the first and second derivative of the strained potential:

\begin{equation}
\frac{\partial U}{\partial \epsilon_{\alpha\beta}}; \quad\quad \frac{\partial^2 U}{\partial\epsilon_{\alpha\beta}\partial\epsilon_{\eta\nu}}.
\end{equation}

Let us recall that for the polymer case the potential energy is given by:

\begin{equation}
U=\sum\limits_{\langle ij \rangle}^{N}\phi_{\rm LJ}^{ij} +
    \sum\limits_{k=1}^{N_p}\sum\limits_{i=1}^{n-1}\chi_{k}^{i} +
    \sum\limits_{k=1}^{N_p}\sum\limits_{i=2}^{n-1}\psi^{i}_{k}
\end{equation}

Regarding the pair-wise interactions we have:

\begin{equation}
\frac{\partial U_{\rm LJ\ or\ FENE}}{\partial \epsilon_{\alpha\beta}}=\frac{\partial \phi_{\rm LJ}^{ij}}{\partial r^{ij}}\frac{\partial r^{ij}}{\partial \epsilon_{\alpha\beta}}\quad \text{or,}~~\frac{\partial \chi^{ij}}{\partial r^{ij}}\frac{\partial r^{ij}}{\partial \epsilon_{\alpha\beta}}
\end{equation}

In order to compute $\partial r_{ij}/\partial \epsilon_{\alpha\beta}$ we define the change in $r^{ij}$ using $\hat{r}_\alpha^{ij}=h_{\alpha\beta}r_\beta^{ij}$. Therefore,

\begin{eqnarray}
\hat{r}^{ij}&=&\sqrt{(\hat{r}_\lambda^{ij})^2} \\
&=&\sqrt{r_\alpha^{ij}h^Thr_\beta^{ij}} \\
&\approx&\sqrt{(r^{ij})^2+2\epsilon_{\alpha\beta}r_\alpha^{ij} r_\beta^{ij}} \\
&=&r^{ij}\sqrt{1+\frac{2\epsilon_{\alpha\beta}r_\alpha^{ij} r_\beta^{ij}}{(r^{ij})^2}} \\
&\approx&r^{ij}\left(1+\frac{\epsilon_{\alpha\beta}r_\alpha^{ij} r_\beta^{ij}}{(r^{ij})^2}-\frac{1}{2}\frac{(\epsilon_{\alpha\beta}r_\alpha^{ij}r_\beta^{ij})^2}{(r^{ij})^4}+\mathcal O(\epsilon^3)\right) \\
&=&r^{ij}+\frac{\epsilon_{\alpha\beta}r_\alpha^{ij} r_\beta^{ij}}{r^{ij}}-\frac{1}{2}\frac{(\epsilon_{\alpha\beta}r_\alpha^{ij}r_\beta^{ij})^2}{(r^{ij})^3}+\mathcal O(\epsilon^3).
\end{eqnarray}

Considering the coefficient of first order term we get:

\begin{equation}
\frac{\partial r_{ij}}{\partial \epsilon_{\alpha\beta}}=\frac{r_\alpha^{ij} r_\beta^{ij}}{r^{ij}},
\end{equation}

and from the the second order term we have:
\begin{eqnarray}
\frac{\partial^2 U_{\rm LJ}}{\partial \epsilon_{\eta\nu}\partial \epsilon_{\alpha\beta}}&=&\frac{\partial}{\partial\epsilon_{\eta\nu}}\left(\frac{\partial \phi^{ij}}{\partial r_{ij}}\frac{\partial r_{ij}}{\partial \epsilon_{\alpha\beta}}\right) \\
&=&\frac{\partial^2 \phi^{ij}}{\partial (r_{ij})^2}\frac{\partial r_{ij}}{\partial \epsilon_{\eta\nu}}\frac{\partial r_{ij}}{\partial \epsilon_{\alpha\beta}}+\frac{\partial \phi^{ij}}{\partial r_{ij}}\frac{\partial^2 r_{ij}}{\partial \epsilon_{\eta\nu}\partial \epsilon_{\alpha\beta}}.
\end{eqnarray}

Using the aforementioned definition of the change in $r^{ij}$ we obtain:
\begin{equation}
\frac{\partial^2 r_{ij}}{\partial \epsilon_{\eta\nu}\partial \epsilon_{\alpha\beta}}=2\frac{(r_\alpha^{ij})^2 (r_\beta^{ij})^2}{(r^{ij})^3}.
\end{equation}

We now turn for the computation of the contribution in the shear modulus coming from the angular part of the potential $\psi$, which is:
\begin{equation}
\frac{\partial \psi^\ell}{\partial \epsilon_{\alpha\beta}}=\frac{\partial \psi^{\ell}}{\partial \cos\varphi^\ell}\frac{\partial \cos\varphi^\ell}{\partial \epsilon_{\alpha\beta}},
\end{equation}
\begin{eqnarray}
\frac{\partial^2 \psi^\ell}{\partial \epsilon_{\eta\nu}\partial \epsilon_{\alpha\beta}}&=&\frac{\partial}{\partial \epsilon_{\eta\nu}}\left(\frac{\partial \psi^{\ell}}{\partial \cos\varphi^\ell}\frac{\partial \cos\varphi^\ell}{\partial \epsilon_{\alpha\beta}}\right)
=\frac{\partial^2 \psi^{\ell}}{\partial (\cos\varphi^\ell)^2}\frac{\partial \cos\varphi^\ell}{\partial \epsilon_{\eta\nu}} \nonumber \\
&&\frac{\partial \cos\varphi^\ell}{\partial \epsilon_{\alpha\beta}}
+\frac{\partial \psi^{\ell}}{\partial \cos\varphi^\ell}\frac{\partial^2 \cos\varphi^\ell}{\partial \epsilon_{\eta\nu}\partial \epsilon_{\alpha\beta}}.
\end{eqnarray}

The terms that remained to computed are :

\begin{equation}
\frac{\partial \cos\varphi^\ell}{\partial \epsilon_{\alpha\beta}}; \quad\quad \frac{\partial^2 \cos\varphi^\ell}{\partial\epsilon_{\alpha\beta}\partial\epsilon_{\eta\nu}}
\end{equation}

Using the definition of the cosine as:

\[
\cos \varphi^{l}=-\frac{r_{\gamma }^{l-1,l} r_{\gamma }^{l,l+1}
}{r^{l-1,l} r^{l,l+1} },r_{\gamma }^{kl} =x_{\gamma }^{l} -x_{\gamma
}^{k},
\]

we have
\begin{widetext}
\[
\begin{array}{l}
 \frac{\partial \cos \varphi^{l}}{\partial \epsilon_{\alpha\beta}}=-\left[
{\frac{r_{\gamma }^{l,l+1} }{r^{l,l+1} }\frac{\partial }{\partial
\epsilon_{\alpha\beta}}\left( {\frac{r_{\gamma }^{l-1,l} }{r^{l-1,l} }}
\right)+\frac{r_{\gamma }^{l-1,l} }{r^{l-1,l} }\frac{\partial }{\partial
\epsilon_{\alpha\beta}}\left( {\frac{r_{\gamma }^{l,l+1} }{r^{l,l+1} }}
\right)} \right] \\
 \frac{\partial^{2}\cos \varphi^{l}}{\partial \epsilon_{\eta\nu} \partial
\epsilon_{\alpha\beta} }=-\left[ {\begin{array}{l}
 \frac{r_{\gamma }^{l,l+1} }{r^{l,l+1} }\frac{\partial^{2}}{\partial
\epsilon_{\eta\nu}\partial\epsilon_{\alpha\beta} }\left( {\frac{r_{\gamma }^{l-1,l}
}{r^{l-1,l} }} \right)+\frac{r_{\gamma }^{l-1,l} }{r^{l-1,l}
}\frac{\partial^{2}}{\partial \epsilon_{\eta\nu} \partial \epsilon_{\alpha\beta}
}\left( {\frac{r_{\gamma }^{l,l+1} }{r^{l,l+1} }} \right)+ \\
 \frac{\partial }{\partial \epsilon_{\alpha\beta} }\left( {\frac{r_{\gamma
}^{l-1,l} }{r^{l-1,l} }} \right)\frac{\partial }{\partial \epsilon_{\eta\nu}
}\left( {\frac{r_{\gamma }^{l,l+1} }{r^{l,l+1} }} \right)+\frac{\partial
}{\partial \epsilon_{\eta\nu} }\left( {\frac{r_{\gamma }^{l-1,l} }{r^{l-1,l}
}} \right)\frac{\partial }{\partial \epsilon_{\alpha\beta} }\left( {\frac{r_{\gamma
}^{l,l+1} }{r^{l,l+1} }} \right) \\
 \end{array}} \right] \\
 \end{array}
\]

where:
\[
\begin{array}{l}
\frac{\partial}{\partial\epsilon_{\alpha\beta}} \left(\frac{r_{\gamma }^{k\ell}}{r^{k\ell}}\right)=
    \frac{1}{{r^{k\ell}}}\frac{\partial r^{k\ell}_\gamma}{\partial\epsilon_{\alpha\beta}}-\frac{r_{\gamma}^{k\ell}}{(r^{k\ell})^2}
    \frac{\partial r^{k\ell}}{\partial\epsilon_{\alpha\beta}} \\
\frac{\partial^2}{\partial\epsilon_{\eta\nu}\partial\epsilon_{\alpha\beta}}\left(\frac{r_\gamma^{k\ell}}{r^{k\ell}}\right)=
        -\frac{1}{(r^{k\ell})^2}\frac{\partial r^{k\ell}}{\partial\epsilon_{\eta\nu}}\frac{\partial r^{k\ell}_\gamma}{\partial\epsilon_{\alpha\beta}}
          +\frac{1}{r^{k\ell}}\frac{\partial^2 r^{k\ell}_\gamma}{\partial\epsilon_{\eta\nu}\partial\epsilon_{\alpha\beta}}
          -\frac{1}{(r^{k\ell})^2}\frac{\partial r^{k\ell}}{\partial\epsilon_{\alpha\beta}}\frac{\partial r^{k\ell}_\gamma}{\partial\epsilon_{\eta\nu}}
          +\frac{2r^{k\ell}_\gamma}{(r^{k\ell})^3}\frac{\partial r^{k\ell}}{\partial\epsilon_{\eta\nu}}\frac{\partial r^{k\ell}}{\partial\epsilon_{\alpha\beta}}
          -\frac{r^{k\ell}_\gamma}{(r^{k\ell})^2}\frac{\partial^2 r^{k\ell}}{\partial\epsilon_{\eta\nu}\partial\epsilon_{\alpha\beta}}
           \\
 \end{array}
\]
\end{widetext}
New derivatives that need to be defined are:

\begin{eqnarray}
\frac{\partial r^{k\ell}_x}{\partial \epsilon_{\alpha\beta}}&=&
\begin{cases}
0 &; \alpha=x, \beta=x \\
r^{k\ell}_y &; \alpha=x, \beta=y \\
r^{k\ell}_y &; \alpha=y, \beta=x \\
\approx 0 &; \alpha=y, \beta=y
\end{cases}
\\
\frac{\partial r^{k\ell}_y}{\partial \epsilon_{\alpha\beta}}&=&0 \\
\frac{\partial^2 r^{k\ell}_\gamma}{\partial \epsilon_{\eta\nu}\partial \epsilon_{\alpha\beta}}&=& 0
\end{eqnarray}

Plugging the latter into cosine derivations we have:

\begin{eqnarray}
&&\frac{\partial \cos\varphi^\ell}{\partial \epsilon_{yy}}=
    -\left[
        \frac{r_\gamma^{\ell,\ell+1}}{r^{\ell,\ell+1}}\frac{\partial}{\partial\epsilon_{yy}}\left(\frac{r_{\gamma}^{\ell-1,\ell}}{r^{\ell-1,\ell}}\right)+
        \frac{r_\gamma^{\ell-1,\ell}}{r^{\ell-1,\ell}}\frac{\partial}{\partial\epsilon_{yy}}\left(\frac{r_{\gamma}^{\ell,\ell+1}}{r^{\ell,\ell+1}}\right)
    \right] \nonumber\\
    &&=
    -\left[
        \frac{r_x^{\ell,\ell+1}}{r^{\ell,\ell+1}}\frac{\partial}{\partial\epsilon_{yy}}\left(\frac{r_{x}^{\ell-1,\ell}}{r^{\ell-1,\ell}}\right)+
        \frac{r_y^{\ell,\ell+1}}{r^{\ell,\ell+1}}\frac{\partial}{\partial\epsilon_{yy}}\left(\frac{r_{y}^{\ell-1,\ell}}{r^{\ell-1,\ell}}\right)+ \right.\nonumber \\
&&\left.\quad\frac{r_x^{\ell-1,\ell}}{r^{\ell-1,\ell}}\frac{\partial}{\partial\epsilon_{yy}}\left(\frac{r_{x}^{\ell,\ell+1}}{r^{\ell,\ell+1}}\right)+
        \frac{r_y^{\ell-1,\ell}}{r^{\ell-1,\ell}}\frac{\partial}{\partial\epsilon_{yy}}\left(\frac{r_{y}^{\ell,\ell+1}}{r^{\ell,\ell+1}}\right)
    \right] \nonumber \\
    &&
    -\left[
        \frac{r_x^{\ell,\ell+1}}{r^{\ell,\ell+1}}\left(-\frac{r_{x}^{\ell-1,\ell}(r_{y}^{\ell-1,\ell})^2}{(r^{\ell-1,\ell})^3}\right)+
        \frac{r_y^{\ell,\ell+1}}{r^{\ell,\ell+1}}\left(-\frac{(r_{y}^{\ell-1,\ell})^3}{(r^{\ell-1,\ell})^3}\right)+ \right.\nonumber \\
&&\left.\quad\frac{r_x^{\ell-1,\ell}}{r^{\ell-1,\ell}}\left(-\frac{r_{x}^{\ell,\ell+1}(r_{y}^{\ell,\ell+1})^2}{(r^{\ell,\ell+1})^3}\right)+
        \frac{r_y^{\ell-1,\ell}}{r^{\ell-1,\ell}}\left(-\frac{(r_{y}^{\ell,\ell+1})^3}{(r^{\ell,\ell+1})^3}\right)
    \right]
\end{eqnarray}
\begin{widetext}
\begin{eqnarray}
\frac{\partial \cos\varphi^\ell}{\partial \epsilon_{xy}}&=&
    -\left[
        \frac{r_\gamma^{\ell,\ell+1}}{r^{\ell,\ell+1}}\frac{\partial}{\partial\epsilon_{xy}}\left(\frac{r_{\gamma}^{\ell-1,\ell}}{r^{\ell-1,\ell}}\right)+
        \frac{r_\gamma^{\ell-1,\ell}}{r^{\ell-1,\ell}}\frac{\partial}{\partial\epsilon_{xy}}\left(\frac{r_{\gamma}^{\ell,\ell+1}}{r^{\ell,\ell+1}}\right)
    \right] \\
    &=&
    -\left[
        \frac{r_x^{\ell,\ell+1}}{r^{\ell,\ell+1}}\frac{\partial}{\partial\epsilon_{xy}}\left(\frac{r_{x}^{\ell-1,\ell}}{r^{\ell-1,\ell}}\right)+
        \frac{r_y^{\ell,\ell+1}}{r^{\ell,\ell+1}}\frac{\partial}{\partial\epsilon_{xy}}\left(\frac{r_{y}^{\ell-1,\ell}}{r^{\ell-1,\ell}}\right)+ \right.\nonumber \\
&\quad&\left.\quad\frac{r_x^{\ell-1,\ell}}{r^{\ell-1,\ell}}\frac{\partial}{\partial\epsilon_{xy}}\left(\frac{r_{x}^{\ell,\ell+1}}{r^{\ell,\ell+1}}\right)+
        \frac{r_y^{\ell-1,\ell}}{r^{\ell-1,\ell}}\frac{\partial}{\partial\epsilon_{xy}}\left(\frac{r_{y}^{\ell,\ell+1}}{r^{\ell,\ell+1}}\right)
    \right] \\
    &=&
    -\left[
        \frac{r_x^{\ell,\ell+1}}{r^{\ell,\ell+1}}\left(\frac{2r_y^{\ell-1,\ell}}{r^{\ell-1,\ell}}-\frac{(r_{x}^{\ell-1,\ell})^2r_{y}^{\ell-1,\ell}}{(r^{\ell-1,\ell})^3}\right)+
        \frac{r_y^{\ell,\ell+1}}{r^{\ell,\ell+1}}\left(-\frac{r_{x}^{\ell-1,\ell}(r_{y}^{\ell-1,\ell})^2}{(r^{\ell-1,\ell})^3}\right)+ \right.\nonumber \\
&\quad&\left.\quad\frac{r_x^{\ell-1,\ell}}{r^{\ell-1,\ell}}\left(\frac{2r_y^{\ell,\ell+1}}{r^{\ell,\ell+1}}-\frac{(r_{x}^{\ell,\ell+1})^2r_{y}^{\ell,\ell+1}}{(r^{\ell,\ell+1})^3}\right)+
        \frac{r_y^{\ell-1,\ell}}{r^{\ell-1,\ell}}\left(-\frac{r_{x}^{\ell,\ell+1}(r_{y}^{\ell,\ell+1})^2}{(r^{\ell,\ell+1})^3}\right)
    \right]
\end{eqnarray}

\begin{eqnarray}
\frac{\partial^2\cos\varphi^\ell}{\partial\epsilon_{xy}\partial\epsilon_{xy}}&=&
-\left[
    \frac{r_\gamma^{\ell,\ell+1}}{r^{\ell,\ell+1}}\frac{\partial^2}{\partial\epsilon_{xy}\partial\epsilon_{xy}}\left(\frac{r_\gamma^{\ell-1,\ell}}{r^{\ell-1,\ell}}\right)+
    \frac{r_\gamma^{\ell-1,\ell}}{r^{\ell-1,\ell}}\frac{\partial^2}{\partial\epsilon_{xy}\partial\epsilon_{xy}}\left(\frac{r_\gamma^{\ell,\ell+1}}{r^{\ell,\ell+1}}\right)+ \right.\nonumber\\
&\quad&\left.\quad2\frac{\partial}{\partial\epsilon_{xy}}\left(\frac{r_\gamma^{\ell-1,\ell}}{r^{\ell-1,\ell}}\right)\frac{\partial}{\partial\epsilon_{xy}}\left(\frac{r_\gamma^{\ell,\ell+1}}{r^{\ell,\ell+1}}\right)
\right]\\
&=&-\left[
    \frac{r_x^{\ell,\ell+1}}{r^{\ell,\ell+1}}\frac{\partial^2}{\partial\epsilon_{xy}\partial\epsilon_{xy}}\left(\frac{r_x^{\ell-1,\ell}}{r^{\ell-1,\ell}}\right)+
    \frac{r_y^{\ell,\ell+1}}{r^{\ell,\ell+1}}\frac{\partial^2}{\partial\epsilon_{xy}\partial\epsilon_{xy}}\left(\frac{r_y^{\ell-1,\ell}}{r^{\ell-1,\ell}}\right)+\right.\nonumber\\
\quad&\quad&\frac{r_x^{\ell-1,\ell}}{r^{\ell-1,\ell}}\frac{\partial^2}{\partial\epsilon_{xy}\partial\epsilon_{xy}}\left(\frac{r_x^{\ell,\ell+1}}{r^{\ell,\ell+1}}\right)+
       \frac{r_y^{\ell-1,\ell}}{r^{\ell-1,\ell}}\frac{\partial^2}{\partial\epsilon_{xy}\partial\epsilon_{xy}}\left(\frac{r_y^{\ell,\ell+1}}{r^{\ell,\ell+1}}\right)+ \nonumber\\
\quad&\quad&\left.2\frac{\partial}{\partial\epsilon_{xy}}\left(\frac{r_x^{\ell-1,\ell}}{r^{\ell-1,\ell}}\right)\frac{\partial}{\partial\epsilon_{xy}}\left(\frac{r_x^{\ell,\ell+1}}{r^{\ell,\ell+1}}\right)+
    2\frac{\partial}{\partial\epsilon_{xy}}\left(\frac{r_y^{\ell-1,\ell}}{r^{\ell-1,\ell}}\right)\frac{\partial}{\partial\epsilon_{xy}}\left(\frac{r_y^{\ell,\ell+1}}{r^{\ell,\ell+1}}\right)
\right]\\
&=&-\left[
    \frac{r_x^{\ell,\ell+1}}{r^{\ell,\ell+1}}\left(
        -2\frac{1}{(r^{\ell-1,\ell})^2}\frac{\partial r^{\ell-1,\ell}}{\partial\epsilon_{xy}}\frac{\partial r^{\ell-1,\ell}_x}{\partial\epsilon_{xy}}
        +2\frac{r^{\ell-1,\ell}_x}{(r^{\ell-1,\ell})^3}\frac{\partial r^{\ell-1,\ell}}{\partial\epsilon_{xy}}\frac{\partial r^{\ell-1,\ell}}{\partial\epsilon_{xy}}
    \right)+
    \frac{r_y^{\ell,\ell+1}}{r^{\ell,\ell+1}}\left(
        2\frac{r^{\ell-1,\ell}_y}{(r^{\ell-1,\ell})^3}\frac{\partial r^{\ell-1,\ell}}{\partial\epsilon_{xy}}\frac{\partial r^{\ell-1,\ell}}{\partial\epsilon_{xy}}
    \right)+\right.\nonumber\\
\quad&\quad&\frac{r_x^{\ell-1,\ell}}{r^{\ell-1,\ell}}\left(
        -2\frac{1}{(r^{\ell,\ell+1})^2}\frac{\partial r^{\ell,\ell+1}}{\partial\epsilon_{xy}}\frac{\partial r^{\ell,\ell+1}_x}{\partial\epsilon_{xy}}
        +2\frac{r^{\ell,\ell+1}_x}{(r^{\ell,\ell+1})^3}\frac{\partial r^{\ell,\ell+1}}{\partial\epsilon_{xy}}\frac{\partial r^{\ell,\ell+1}}{\partial\epsilon_{xy}}
    \right)+
       \frac{r_y^{\ell-1,\ell}}{r^{\ell-1,\ell}}\left(
        2\frac{r^{\ell,\ell+1}_y}{(r^{\ell,\ell+1})^3}\frac{\partial r^{\ell,\ell+1}}{\partial\epsilon_{xy}}\frac{\partial r^{\ell,\ell+1}}{\partial\epsilon_{xy}}
    \right)+ \nonumber\\
\quad&\quad&        2\left(\frac{1}{{r^{\ell-1,\ell}}}\frac{\partial r^{\ell-1,\ell}_x}{\partial\epsilon_{xy}}-\frac{r_x^{\ell-1,\ell}}{(r^{\ell-1,\ell})^2}
    \frac{\partial r^{\ell-1,\ell}}{\partial\epsilon_{xy}}\right)
                     \left(\frac{1}{{r^{\ell,\ell+1}}}\frac{\partial r^{\ell,\ell+1}_x}{\partial\epsilon_{xy}}-\frac{r_x^{\ell,\ell+1}}{(r^{\ell,\ell+1})^2}
    \frac{\partial r^{\ell,\ell+1}}{\partial\epsilon_{xy}}\right)+\nonumber\\
\quad&\quad&\left. 2\left(\frac{1}{{r^{\ell-1,\ell}}}\frac{\partial r^{\ell-1,\ell}_y}{\partial\epsilon_{xy}}-\frac{r_y^{\ell-1,\ell}}{(r^{\ell-1,\ell})^2}
    \frac{\partial r^{\ell-1,\ell}}{\partial\epsilon_{xy}}\right)
                     \left(\frac{1}{{r^{\ell,\ell+1}}}\frac{\partial r^{\ell,\ell+1}_y}{\partial\epsilon_{xy}}-\frac{r_y^{\ell,\ell+1}}{(r^{\ell,\ell+1})^2}
    \frac{\partial r^{\ell,\ell+1}}{\partial\epsilon_{xy}}\right)
\right]\\
&=&-\left[
    \frac{r_x^{\ell,\ell+1}}{r^{\ell,\ell+1}}\left(
        -\frac{4r^{\ell-1,\ell}_x(r^{\ell-1,\ell}_y)^2}{(r^{\ell-1,\ell})^4}
        +\frac{2(r^{\ell-1,\ell}_x)^3(r^{\ell-1,\ell}_y)^2}{(r^{\ell-1,\ell})^5}
    \right)+
    \frac{r_y^{\ell,\ell+1}}{r^{\ell,\ell+1}}\left(
        \frac{2(r^{\ell-1,\ell}_x)^2(r^{\ell-1,\ell}_y)^3}{(r^{\ell-1,\ell})^5}
    \right)+\right.\nonumber\\
\quad&\quad&\frac{r_x^{\ell-1,\ell}}{r^{\ell-1,\ell}}\left(
        -\frac{4r^{\ell,\ell+1}_x(r^{\ell,\ell+1}_y)^2}{(r^{\ell,\ell+1})^4}
        +\frac{2(r^{\ell,\ell+1}_x)^3(r^{\ell,\ell+1}_y)^2}{(r^{\ell,\ell+1})^5}
    \right)+
       \frac{r_y^{\ell-1,\ell}}{r^{\ell-1,\ell}}\left(
        \frac{2(r^{\ell,\ell+1}_x)^2(r^{\ell,\ell+1}_y)^3}{(r^{\ell,\ell+1})^5}
    \right)+ \nonumber\\
\quad&\quad&        2\left(\frac{2r^{\ell-1,\ell}_y}{r^{\ell-1,\ell}}-\frac{(r_x^{\ell-1,\ell})^2r_y^{\ell-1,\ell}}{(r^{\ell-1,\ell})^3}
                    \right)
                     \left(\frac{2r^{\ell,\ell+1}_y}{{r^{\ell,\ell+1}}}-\frac{(r_x^{\ell,\ell+1})^2r_y^{\ell,\ell+1}}{(r^{\ell,\ell+1})^3}
                    \right)+\nonumber\\
\quad&\quad&\left. 2\left(-\frac{r_x^{\ell-1,\ell}(r_y^{\ell-1,\ell})^2}{(r^{\ell-1,\ell})^3}
                    \right)
                     \left(-\frac{r_x^{\ell,\ell+1}(r_y^{\ell,\ell+1})^2}{(r^{\ell,\ell+1})^3}
                    \right)
\right]
\end{eqnarray}

For non-affine forces$=0$:
\begin{eqnarray}
\frac{\partial^2 \cos\varphi^\ell}{\partial x_\nu^m \partial \epsilon_{xy}}&=&
    -\frac{\partial}{\partial \epsilon_{xy}}\left[ {\left( {\delta^{\ell,m}-\delta^{\ell-1,m}} \right)\left(
{\frac{r_{\nu }^{\ell,l+1} }{r^{\ell,\ell+1} r^{\ell-1,\ell} }-\frac{r_{\nu
}^{\ell-1,\ell} r_{\gamma }^{\ell-1,\ell} r_{\gamma }^{\ell,\ell+1} }{r^{\ell,\ell+1}
(r^{\ell-1,\ell} )^{3}}} \right)+ }\right. \nonumber \\
\quad&\quad&\quad\quad\quad\left.{\left( {\delta^{\ell+1,m}-\delta^{\ell,m}}
\right)\left( {\frac{r_{\nu }^{\ell-1,\ell} }{r^{\ell,\ell+1} r^{\ell-1,\ell}
}-\frac{r_{\nu }^{\ell,\ell+1} r_{\gamma }^{\ell-1,\ell} r_{\gamma }^{\ell,\ell+1}
}{r^{\ell-1,\ell} (r^{\ell,\ell+1} )^{3}}} \right)} \right]
 \\
 &=&
    {\left( {\delta^{\ell-1,m}-\delta^{\ell,m}} \right)\left(\frac{\partial}{\partial \epsilon_{xy}}
{\frac{r_{\nu }^{\ell,l+1} }{r^{\ell,\ell+1} r^{\ell-1,\ell} }-\frac{\partial}{\partial \epsilon_{xy}}\frac{r_{\nu
}^{\ell-1,\ell} r_{\gamma }^{\ell-1,\ell} r_{\gamma }^{\ell,\ell+1} }{r^{\ell,\ell+1}
(r^{\ell-1,\ell} )^{3}}} \right)+ } \nonumber \\
\quad\quad\quad&\quad&\left.{\left( {\delta^{\ell,m}-\delta^{\ell+1,m}}
\right)\left( {\frac{\partial}{\partial \epsilon_{xy}}\frac{r_{\nu }^{\ell-1,\ell} }{r^{\ell,\ell+1} r^{\ell-1,\ell}
}-\frac{\partial}{\partial \epsilon_{xy}}\frac{r_{\nu }^{\ell,\ell+1} r_{\gamma }^{\ell-1,\ell} r_{\gamma }^{\ell,\ell+1}
}{r^{\ell-1,\ell} (r^{\ell,\ell+1} )^{3}}} \right)} \right]
 \\
 &=&
    \left( {\delta^{\ell-1,m}-\delta^{\ell,m}} \right)\left(\frac{1}{ r^{\ell-1,\ell} }\frac{\partial}{\partial \epsilon_{xy}}
\frac{r_{\nu }^{\ell,l+1}}{r^{\ell,\ell+1}}+\frac{r_{\nu }^{\ell,l+1}}{r^{\ell,\ell+1}}\frac{\partial}{\partial \epsilon_{xy}}\frac{1}{ r^{\ell-1,\ell} }\right.\nonumber\\
&\quad&\quad\quad\quad\quad\quad\quad-\left.\frac{r_{\nu}^{\ell-1,\ell} r_{\gamma }^{\ell-1,\ell}}{(r^{\ell-1,\ell} )^{3}}\frac{\partial}{\partial \epsilon_{xy}}\frac{r_{\gamma}^{\ell,\ell+1}}{r^{\ell,\ell+1}}
 -\frac{r_{\gamma}^{\ell,\ell+1}}{r^{\ell,\ell+1}}\frac{\partial}{\partial \epsilon_{xy}}\frac{r_{\nu}^{\ell-1,\ell} r_{\gamma }^{\ell-1,\ell}}{(r^{\ell-1,\ell} )^{3}}
 \right)+ \nonumber \\
&\quad&\left( {\delta^{\ell,m}-\delta^{\ell+1,m}}\right)\left( \frac{1}{r^{\ell,\ell+1}}\frac{\partial}{\partial \epsilon_{xy}}\frac{r_{\nu }^{\ell-1,\ell} }{ r^{\ell-1,\ell}}+
\frac{r_{\nu }^{\ell-1,\ell} }{ r^{\ell-1,\ell}}\frac{\partial}{\partial \epsilon_{xy}}\frac{1}{r^{\ell,\ell+1}} \right.\nonumber \\
&\quad&\quad\quad\quad\quad\quad\quad\left.\left.-\frac{r_{\nu }^{\ell,\ell+1}  r_{\gamma }^{\ell,\ell+1}
}{(r^{\ell,\ell+1} )^{3}}\frac{\partial}{\partial \epsilon_{xy}}\frac{r_{\gamma }^{\ell-1,\ell}}{r^{\ell-1,\ell}}
-\frac{r_{\gamma }^{\ell-1,\ell}}{r^{\ell-1,\ell}}\frac{\partial}{\partial \epsilon_{xy}}\frac{r_{\nu }^{\ell,\ell+1}  r_{\gamma }^{\ell,\ell+1}
}{(r^{\ell,\ell+1} )^{3}} \right) \right]
\end{eqnarray}
Using previously defined expressions and:
\begin{equation}
\frac{\partial}{\partial\epsilon_{\alpha\beta}}\frac{r_\nu^{k\ell}r_\gamma^{k\ell}}{(r^{k\ell})^3}=
\frac{r_\gamma^{k\ell}}{(r^{k\ell})^3}\frac{\partial r_\nu^{k\ell}}{\partial\epsilon_{\alpha\beta}}+
\frac{r_\nu^{k\ell}}{(r^{k\ell})^3}\frac{\partial r_\gamma^{k\ell}}{\partial\epsilon_{\alpha\beta}}-
\frac{3r_\nu^{k\ell}r_\gamma^{k\ell}}{(r^{k\ell})^4}\frac{\partial r^{k\ell}}{\partial\epsilon_{\alpha\beta}}
\end{equation}
we have the final expression of the second derivative of cosine as:
\begin{eqnarray}
\frac{\partial^2 \cos\varphi^\ell}{\partial x_\nu^m \partial \epsilon_{xy}} &=&
    \left( {\delta^{\ell-1,m}-\delta^{\ell,m}} \right)\left[\frac{1}{ r^{\ell-1,\ell} }
    \left(\frac{\delta^{\nu x}2r_y^{\ell,\ell+1}}{r^{\ell,\ell+1}}-\frac{r_{\nu }^{\ell,\ell+1}r_{x}^{\ell,\ell+1}r_{y}^{\ell,\ell+1}}{(r^{\ell,\ell+1})^3}\right)
-\frac{r_{\nu }^{\ell,l+1}}{r^{\ell,\ell+1}}\left(\frac{r_x^{\ell-1,\ell} r_y^{\ell-1,\ell}}{ (r^{\ell-1,\ell})^3 }\right)\right.\nonumber\\
&\quad&\quad\quad\quad\quad\quad\quad-\frac{r_{\nu}^{\ell-1,\ell} r_{\gamma }^{\ell-1,\ell}}{(r^{\ell-1,\ell} )^{3}}\left(\frac{2r_y^{\ell,l+1}}{r^{\ell,\ell+1}}-\frac{r_{\gamma }^{\ell,l+1}r_{x}^{\ell,l+1}r_{y}^{\ell,l+1}}{(r^{\ell,\ell+1})^3}\right)\nonumber\\
&\quad&\quad\quad\quad\quad\quad\quad\left.-\frac{r_{\gamma}^{\ell,\ell+1}}{r^{\ell,\ell+1}}\left(\frac{2r_y^{\ell-1,\ell}(\delta^{\nu x}r_\gamma^{\ell-1,\ell}+r_\nu^{\ell-1,\ell})}{(r^{\ell-1,\ell})^3}-\frac{3r_\gamma^{\ell-1,\ell}r_\nu^{\ell-1,\ell}r_x^{\ell-1,\ell}r_y^{\ell-1,\ell}}{(r^{\ell-1,\ell})^5}\right)
 \right]+ \nonumber \\
&\quad&\left( {\delta^{\ell,m}-\delta^{\ell+1,m}}\right)\left[ \frac{1}{r^{\ell,\ell+1}}
\left(\frac{\delta^{\nu x}2r_y^{\ell-1,\ell}}{r^{\ell-1,\ell}}-\frac{r_{\nu }^{\ell-1,\ell}r_{x}^{\ell-1,\ell}r_{y}^{\ell-1,\ell}}{(r^{\ell-1,\ell})^3}\right)+
\frac{r_{\nu }^{\ell-1,\ell} }{ r^{\ell-1,\ell}}\left(\frac{r_x^{\ell,\ell+1} r_y^{\ell,\ell+1}}{ (r^{\ell,\ell+1})^3 }\right) \right.\nonumber \\
&\quad&\quad\quad\quad\quad\quad\quad-\frac{r_{\nu }^{\ell,\ell+1}  r_{\gamma }^{\ell,\ell+1}}{(r^{\ell,\ell+1} )^{3}}\left(\frac{2r_y^{\ell-1,\ell}}{r^{\ell-1,\ell}}-\frac{r_{\nu }^{\ell-1,\ell}r_{x}^{\ell-1,\ell}r_{y}^{\ell-1,\ell}}{(r^{\ell-1,\ell})^3}\right) \nonumber \\
&\quad&\quad\quad\quad\quad\quad\quad\left.-\frac{r_{\gamma }^{\ell-1,\ell}}{r^{\ell-1,\ell}}\left(\frac{2r_y^{\ell,\ell+1}(\delta^{\nu x} r_\gamma^{\ell,\ell+1}+ r_\nu^{\ell,\ell+1})}{(r^{\ell,\ell+1})^3}- \frac{3r_\gamma^{\ell,\ell+1}r_\nu^{\ell,\ell+1}r_x^{\ell,\ell+1}r_y^{\ell,\ell+1}}{(r^{\ell,\ell+1})^5}\right) \right]
\end{eqnarray}
\end{widetext}

\end{document}